\documentclass[prd,preprint,superscriptaddress,amsmath,amssymb]{revtex4}

 \pdfoutput=1

\usepackage{graphicx}
\usepackage{color,slashed}

\begin{document}
\title{\bf Top-quark  FCNC decays, LFVs,  lepton $g-2$, and $W$ mass anomaly with inert charged Higgses}

\author{Chuan-Hung Chen}
\email[E-mail: ]{physchen@mail.ncku.edu.tw}
\affiliation{Department of Physics, National Cheng-Kung University, Tainan 70101, Taiwan}
\affiliation{Physics Division, National Center for Theoretical Sciences, Taipei 10617, Taiwan}

\author{Cheng-Wei Chiang}
\email[E-mail: ]{chengwei@phys.ntu.edu.tw}
\affiliation{Department of Physics and Center for Theoretical Physics, National Taiwan University, Taipei 10617, Taiwan}
\affiliation{Physics Division, National Center for Theoretical Sciences, Taipei 10617, Taiwan}

\author{Chun-Wei Su}
\email[E-mail: ]{r10222026@ntu.edu.tw}
\affiliation{Department of Physics and Center for Theoretical Physics, National Taiwan University, Taipei 10617, Taiwan}

\date{\today}

\begin{abstract}

The observed flavor-changing neutral-current (FCNC) processes in the standard model (SM) arise from the loop diagrams involving the weak charged currents mediated by the $W$-gauge boson.  Nevertheless, the top-quark FCNCs and lepton-flavor violating processes resulting from the same mechanism are highly suppressed.  We investigate possible new physics effects that can enhance the suppressed FCNC processes, such as $t\to q(h,V)$ with $V=\gamma,Z,g$, $h\to \ell \ell'$, and $\ell\to \ell' \gamma$. To achieve the assumption that the induced-FCNCs are all from quantum loops, we consider the scotogenic mechanism, where a $Z_2$ symmetry is introduced and only new particles carry an odd $Z_2$ parity. With the extension of the SM to include an inert Higgs doublet, an inert charged Higgs singlet, a vector-like singlet quark, and two neutral leptons, it is found that, with relevant constraints taken into account, the $t\to c (h, Z)$, $h\to \mu\tau$, and $\tau\to \ell \gamma$ decays can be enhanced up to the expected sensitivities in experiments.  The branching ratios of $h\to \mu^+ \mu^-/\tau^+ \tau^-$ from only new physics effects can reach up to ${\cal O}(10^{-3})$. Intriguingly, the resulting muon $g-2$ can fit the combined data within $2\sigma$ errors, whereas the electron $g-2$ can have either sign with a magnitude of ${\cal O}(10^{-13}-10^{-12})$. In addition, we examine the oblique parameters in the model and find that the resulting $W$-mass anomaly observed by CDF II can be accommodated. 

\end{abstract}
\maketitle

\section{Introduction} \label{sec:Introduction}

While flavor-changing processes at tree level in the standard model (SM) arise from weak charged currents mediated by the $W$ gauge boson, flavor-changing neutral currents (FCNCs) occur only via quantum loops and have been observed in various experiments, notably $P-\bar P$ mixing and $P\to P' \ell^+ \ell^-$ decays with $P^{(\prime)}$ being the $K$, $D$, and $B$ mesons.  However, not all loop-induced FCNC processes in the SM are sufficiently sizable and detectable under current experimental sensitivities.  For instance, due to the Glashow-Iliopoulos-Maiani (GIM) mechanism~\cite{Glashow:1970gm}, the top-quark FCNCs are highly suppressed, and the branching ratios (BRs) for the $t\to q (g, \gamma, Z, h)$ decays with $q=u, c$ are of the order  of $10^{-12}-10^{-17}$~\cite{AguilarSaavedra:2004wm,Abbas:2015cua,Balaji:2020qjg}.  A similar suppression also happens in lepton flavor-violating (LFV) processes, e.g., $\ell\to \ell' \gamma$, $\ell\to 3\ell'$, and $h\to \ell \ell'$.

The expected sensitivities in the high-luminosity (HL) LHC with an integrated luminosity of 3ab$^{-1}$ at $\sqrt{s}=14$~TeV are expected to reach ${\cal O}(10^{-5})$ for $t\to q Z/q\gamma$, ${\cal O}(10^{-4})$ for $t\to q h$~\cite{Azzi:2019yne}, and ${\cal O}(10^{-4})$ for $h\to e\tau/\mu\tau$~\cite{Cepeda:2019klc}. In addition, the $\mu \to e \gamma$ decay in MEG II experiment can reach the sensitivity of $6\times 10^{-14}$~\cite{MEGII:2018kmf}, and $\tau \to e\gamma/\mu\gamma$ can be probed at the level of ${\cal O}(10^{-9})$ in Belle II~\cite{Belle-II:2018jsg}. Thus, if any signal in these processes is detected in experiments, it definitely indicates new physics effects at play. Some interesting extensions of the SM proposed to enhance the top-FCNC decays can be found in Refs.~\cite{Abraham:2000kx,Eilam:2001dh,AguilarSaavedra:2002kr,Dey:2016cve,Gaitan:2017tka,Shen:2017oel,Chiang:2018oyd,Oyulmaz:2018irs,Chen:2018lze,Arroyo-Urena:2019qhl,Shi:2019epw,Liu:2020kxt,Hou:2020ciy,Bie:2020sro,Gutierrez:2020eby,Liu:2021crr,Cai:2022xha,Hernandez-Juarez:2022kjx,Badziak:2017wxn,Chen:2022dzc}.

A long-standing anomaly in the muon anomalous magnetic dipole moment (muon $g-2$) observed at BNL~\cite{Muong-2:2006rrc} is now supported by the recent new measurement performed in the E989 Run 1 experiment at Fermilab~\cite{Muong-2:2021ojo}.  The combined data shows a $4.2\sigma$ deviation from the SM prediction, which is obtained by the data-driven evaluations of hadronic vacuum polarization (HVP)~\cite{Aoyama:2020ynm}:
  \begin{equation}
  \Delta a_\mu=a^{\rm exp}_\mu-a^{\rm SM}_\mu = (2.51\pm 0.59)\times 10^{-9}\,. \label{eq:Damu}
  \end{equation}
We note that although the discrepancy in Eq.~(\ref{eq:Damu}) could possibly be narrowed down according to the calculations of  lattice QCD~\cite{FermilabLattice:2019ugu,Borsanyi:2020mff,Ce:2022kxy,Alexandrou:2022amy,Blum:2023qou}, the lattice results lead to a tension with $e^+ e^- \to$ hadrons cross section data~\cite{Crivellin:2020zul,Keshavarzi:2020bfy,Colangelo:2020lcg,Colangelo:2022vok}. Hence, the discrepancy between theoretical estimates and data has not been completely resolved yet. In addition, using precision measurements of the fine structure constant, electron $g-2$ measured separately using cesium~\cite{electron_gm2_Cs} and rubidium~\cite{electron_gm2_Rb} atoms is respectively given by:
  \begin{align}
  \Delta a_e ({\rm Cs})& = (-8.8\pm 3.6) \times 10^{-13}\,, \nonumber \\
  \Delta a_e ({\rm Rb}) &=(4.8 \pm 3.0)\times  10^{-13}\,. \label{eq:Dae}
  \end{align}
Further precision measurement needs to be done in order to resolve the above discrepancy and to tell us whether the data agree with the SM prediction.  In any case, a significant deviation from the SM prediction in lepton $g-2$ is an important channel to probe new physics effects in the lepton sector~\cite{Chen:2001kn,Chen:2017hir,Han:2018znu,Chen:2019nud,Chen:2020ptg,Chen:2020jvl,Chen:2020tfr,Dorsner:2020aaz,Jana:2020joi,Chun:2020uzw,Li:2020dbg,Bodas:2021fsy,Baker:2021yli,Chiang:2021pma,Chen:2021jok,Yang:2021duj,Athron:2021iuf,Escribano:2021css,Cen:2021ryk,Borah:2021jzu,Jueid:2021avn,Dey:2021pyn,Li:2021koa,Hue:2021xzl,Chiang:2022axu,Chowdhury:2022jde,Li:2022zap,Arora:2022hza}.

Using  the full dataset of the integrated luminosity of $8.8$ fb$^{-1}$ in proton-antiproton collisions at $\sqrt{s}=1.96$~TeV, the CDF II Collaboration recently reported the measured mass of $W$ gauge boson as:
  \begin{equation}
  m_W = 80.4335\pm 0.0094~\text{GeV}\,,
  \end{equation}
where the observed value is different from $m_W=80.370\pm 0.019$~GeV measured by ATLAS~\cite{ATLAS:2017rzl} and earlier result of $m_W=80.385\pm 0.015$~GeV that is the measurement of LEP combined with Tevatron~\cite{CDF:2013dpa}. Moreover, the new observation has a $7\sigma$ deviation from the SM prediction $m_W=80.361$~GeV~\cite{Heinemeyer:2013dia}. If the $W$-mass anomaly is confirmed by the measurements at the LHC with more cumulative data, it would be a solid piece of evidence that exhibits the effects of new physics~\cite{Fan:2022dck,Strumia:2022qkt,Bagnaschi:2022whn,Bahl:2022xzi,Cheng:2022jyi,Asadi:2022xiy,Heckman:2022the,Crivellin:2022fdf,FileviezPerez:2022lxp,Kanemura:2022ahw,Kim:2022hvh,Li:2022gwc,Dcruz:2022dao,Chowdhury:2022dps,Gao:2022wxk,Han:2022juu,Cheng:2022hbo,Bandyopadhyay:2022bgx}.


In this work, we investigate a new physics model that can enhance the top-FCNC and LFV processes, up to the experimental sensitivities mentioned above.  Furthermore, we will show that, after all possible constraints being taken into account, these new physics effects can lead to $\Delta a_\mu$ of ${\cal O}(10^{-9})$ and $|\Delta a_e|$ of ${\cal O}(10^{-13}-10^{-12})$ and explain the $W$-mass anomaly.

Note that the $t\to q (h,Z)$ and $h\to \ell \ell'$ decays can in general proceed via tree-level diagrams through mixing.  Such tree-level effects can be naturally suppressed when the new particles are charged under an unbroken symmetry, for which the SM particles remain neutral.  Under the latter scenario, we consider in this study a SM extension where all FCNCs arise only from loop diagrams.  To achieve the purpose, we impose a $Z_2$ symmetry under which the new and SM particles can be classified as $Z_2$-odd and -even, respectively.  From the initial- and final-state particles involved in the above-mentioned processes, one can infer that the new mediating particles running in the loop diagrams should be $Z_2$-odd scalar bosons and $Z_2$-odd fermions.

In order to realize the above inference based on a gauge anomaly-free model, a minimal extension to the SM includes one inert Higgs doublet~\cite{Barbieri:2006dq},  one charged Higgs singlet~\cite{Zee:1980ai}, one $SU(2)_L$-singlet vector-like quark~\cite{Chen:2022dzc}, and two singlet vector-like Dirac-type neutral leptons~\cite{Chen:2022gmk}, where the $Z_2$-odd quark is responsible for the top-FCNC processes, and the $Z_2$-odd singlet leptons are for the rare lepton flavor-conserving and -violating processes. The new singlet $Z_2$-odd charged Higgs boson can be used to enhance the rare decays and avoid the chirality suppression of $m_F/\Lambda$, where $m_F$ is the mass of the SM particle and $\Lambda$ is the mass scale of heavy particle in the model.

There are three scalar bosons in the inert Higgs doublet, namely, an inert charged Higgs, a scalar, and a pesudoscalar, with the lightest neutral inert scalar being a possible DM candidate.  It is found that the top-FCNC and LFV processes are dominated by the charged Higgs boson. Since fermions of different chiralities couple to different charged Higgs bosons in the model, the chirality-flipping processes $t\to q (h,\gamma)$, $h\to \ell \ell'$, and $\ell\to \ell' \gamma$ strongly depend on the mixing of the two charged Higgs bosons.  Furthermore, because of the charged Higgs mixing, the induced lepton $g-2$ can be either positive or negative, where the contribution from a single charged Higgs boson is usually negative at the one-loop level.


This paper is organized as follows: We introduce the model and derive the relevant gauge couplings, masses of inert scalars, charged Higgs mixing, and Yukawa couplings in Sec.~\ref{sec:model}. The induced top-FCNC and LFV processes including lepton $g-2$ are studied in detail in Sec.~\ref{sec:phenomenology}. In Sec.~\ref{sec:constraints}, we discuss the strict constraints from $\Delta B(D)=2$ transitions, the Higgs to diphoton decay, and the $\mu\to e\gamma$ decay.  In Sec.~\ref{sec:Num}, we comprehensively scan the parameter space by taking into account all major constraints.  Finally, a summary of the work is given in Sec.~\ref{sec:sum}.

\section{Model,  gauge couplings, and Yukawa couplings} \label{sec:model}

To enhance the suppressed FCNC processes in the SM and to explain the muon $g-2$ anomaly and $W$-mass excesses through loop effects, we extend the SM by including one inert Higgs doublet ($H_I$), one charged scalar singlet ($\chi^\pm$), two vector-like lepton singlet  ($N_{1,2}$), and one vector-like quark singlet ($B$) under $SU(2)_L$. In order to obtain a stable dark matter (DM) candidate, we impose a $Z_2$ symmetry in such a way that the new particles are $Z_2$-odd and the SM particles are $Z_2$-even. The representations and charge assignments of $Z_2$-odd particles are given in Table~\ref{tab:rep}, where we use the convention that the electric charge of a field $Q = I_3 + \frac{Y}{2}$ with $I_3$ and $Y$ being the isospin and hypercharge quantum numbers, respectively. Accordingly, we discuss the relevant couplings from the scalar potential, gauge sector, and Yukawa sector in the following subsections. 

\begin{table}[thp]
 \caption{ Representations and charge assignments of new fields. }
\begin{center}
\begin{tabular}{cccccc} \hline \hline
  & ~~$SU(2)_L$~~ & ~~$U(1)_Y$~~  &  ~~$Z_2$~~ & ~~Lepton  \\ \hline
 $H_I$ & 2 & 1 & $-1$ & 0 \\ \hline
 $\chi^+$ &  1 & 2 & $-1$ & 0 \\ \hline 
 $N_{1,2}$ & 1 & 0 & $-1$ & 1 \\ \hline
 $B$ & 1 & $-2/3$ &  $-1$ & 0 \\ \hline \hline
\end{tabular}
\end{center}
\label{tab:rep}
\end{table}%

\subsection{Inert charged Higgs mixing and trilinear Higgs couplings}

With the addition of one inert Higgs doublet and one charged Higgs singlet into the SM, the most general scalar potential consistent with the required symmetries is given by:
 \begin{align}
 V 
 = & 
 \mu^2_1 H^+ H + \mu^2_2 H^\dagger_I H_I + \lambda_1 (H^\dagger H)^2 + \lambda_2 (H^\dagger_I H_I)^2  + \lambda_3 (H^\dagger H) (H^\dagger_I H_I) \nonumber \\
 & + \lambda_4 (H^\dagger H_I) (H^\dagger_I H) + \frac{1}{2} [\lambda_5 (H^\dagger H_I) + H.c.] + m^2_{\chi^\pm} \chi^- \chi^+ \nonumber \\
 & +[\mu_\chi H^T_I i\tau_2 H \chi^- + H.c.]+ \lambda^{\chi}_1 (\chi^- \chi^+)^2 + \lambda^{\chi}_2 H^\dagger H \chi^- \chi^+ + \lambda^\chi_3 H^\dagger_I H_I \chi^- \chi^+  \,, \label{eq:scalar_P}
 \end{align}
where $H$ is the SM Higgs doublet, $\mu_\chi$ has the dimension of mass, and  $\tau_2$ is the second Pauli matrix.   To examine the spectra of scalar bosons, we write the components of the  scalar doublets as:
\begin{equation}
  H= 
\left(
\begin{array}{c}
  G^+     \\
  \frac{1}{\sqrt{2}} ( v + h + i G^0)      \\   
\end{array}
\right)~, ~~  H_I= 
\left(
\begin{array}{c}
  H_I^+    \\
  \frac{1}{\sqrt{2}} (S_I + i A_I)    \\   
\end{array}
\right)~, \label{eq:H_H_I_rep}
\end{equation}
where $\mu^2_1<0$ is assumed to facilitate spontaneous electroweak symmetry breaking (EWSB), $G^{\pm,0}$ are the Goldstone bosons, $v$ is the vacuum expectation value (VEV) of $H$, and  $h$ is the SM Higgs boson.  Note that the components in these two doublet fields do not mix due to the imposed $Z_2$ symmetry.  Moreover, we take $\mu^2_2, m^2_\chi >0$ so that $S_I~ (A_I)$, $H^\pm_I$, and $\chi^\pm$ are massive particles before EWSB.

In addition to the tadpole conditions, i.e., $\partial V/\partial h(S_I) =0$, the vacuum stability is controlled by the co-positivity criteria of the dimension-4 terms in Eq.~(\ref{eq:scalar_P}) and lead to~\cite{Klimenko:1984qx,Kannike:2012pe,Longas:2015sxk}:
\begin{align}
& \lambda_{1,2}, \lambda^\chi_{1,2,3} \geq 0\,, ~ \lambda^{\chi}_{2} + 2\sqrt{\lambda_1 \lambda^\chi_1} >0\,,~ \lambda^{\chi}_{3} + 2\sqrt{\lambda_2 \lambda^\chi_1} >0\,, 
\nonumber \\
& \lambda_3 + 2 \sqrt{\lambda_1 \lambda_2} \geq 0\,, ~ \lambda_3 + \lambda_4 - |\lambda_5| +  2 \sqrt{\lambda_1 \lambda_2} \geq 0\,. \label{eq:BFB}
\end{align}
On the other hand, the tree-level perturbative unitarity of scalar scattering amplitudes requires that $|\lambda_i|, |\lambda^\chi_i| \leq 4 \pi$~\cite{Lee:1977eg}.

With the parametrization in Eq.~(\ref{eq:H_H_I_rep}), the masses of $S_I$ and $A_I$ are given by:
\begin{equation}
 m^2_{S_I} = \mu^2_2 + \frac{\lambda_L v^2}{2}\,, ~m^2_{A_I} = \mu^2_2 + \frac{\lambda_A v^2}{2}\,,
\end{equation}
with $\lambda_{L(A)} \equiv \lambda_3 + \lambda_4 \pm \lambda_5$.  The two $Z_2$-odd charged Higgs bosons in the model can mix through the dimension-3 term $\mu_\chi H^T_I i\tau_2 H \chi^-$ and, therefore, has the mass-square matrix
\begin{align}
& (H^-_I, \chi^-)   
\left(
\begin{array}{cc}
 m^2_{11} &   m^2_{12}   \\
m^2_{12}  &  m^2_{22}    \\    
\end{array}
\right)  \left(
\begin{array}{c}
 H^+_I    \\
  \chi^+      \\    
\end{array}
\right)\,, \nonumber \\
& \mbox{with}~
m^2_{11} = \mu^2_2 + \frac{\lambda_3 v^2}{2}\,, ~m^2_{12} = \frac{\mu_\chi v}{\sqrt{2}}\,, ~ m^2_{22}  = \mu^2_\chi + \frac{\lambda^\chi_2 v^2}{2}\,.
\end{align}
Suppose this $2\times 2$ real symmetric mass-square matrix is diagonalized by an $S{\cal O}(2)$ orthogonal rotation defined by
\begin{equation}
\left(
\begin{array}{c}
 H^+_1  \\
H^+_2      \\    
\end{array}
\right) = \left( \begin{array}{cc}
c_{\theta_\chi}  & c_{\theta_\chi} \\
-s_{\theta_\chi}     &  c_{\theta_\chi} \\    
\end{array}
\right)  \left(
\begin{array}{c}
 H^+_I    \\
 \chi^+      \\    
\end{array}
\right)\,  \label{eq:mixing}
\end{equation}
where $c_{\theta_\chi} \equiv \cos\theta_\chi$ and $s_{\theta_\chi} \equiv \sin\theta_\chi$.
Then we obtain the mass eigenvalues and the mixing angle as:
\begin{align}
  m^2_{1(2)} &= \frac{m^2_{11} + m^2_{22}}{2} \pm \frac{1}{2} \sqrt{(m^2_{22} - m^2_{11})^2 + 4 (m^2_{12})^2}\,, \nonumber \\
  s_{2\theta_\chi} & = - \frac{2 m^2_{12} }{m^2_{2} - m^2_{1} }\,,  \label{eq:theta_chi}
\end{align}
where $s_{2\theta_\chi} \equiv \sin2\theta_\chi$. 
  
Since the $t\to q h$ and $h\to \ell \ell'$ processes involve the Higgs couplings to the $Z_2$-odd scalars, we need to extract the trilinear Higgs couplings to $H^\pm_{1,2}$, $S_I$, and $A_I$ from the scalar potential. According to Eqs.~(\ref{eq:scalar_P}) and (\ref{eq:mixing}), the $h$-$H^-_i$-$H^+_i$ and $h$-$S_I(A_I)$-$S_I(A_I)$ interactions can be written as:
\begin{align}
{\cal L}_{h\phi_i \phi_j}  & =-   \lambda^h_{ij} v \, h H^-_i H^+_j - \frac{1}{2} \lambda_{L} v \, h( S^2_I + A^2_I)\,, \nonumber \\
  \lambda^h_{11} & = \lambda_3 c^2_{\theta_\chi} + \lambda^\chi_2 s^2_{\theta_\chi} -\frac{m^2_{H^\pm_2} - m^2_{H^\pm_1}}{2v^2} s^2_{2\theta_\chi}\,, \nonumber \\
\lambda^h_{22} & = \lambda_3 s^2_{\theta_\chi} + \lambda^\chi_2 c^2_{\theta_\chi} +\frac{m^2_{H^\pm_2} - m^2_{H^\pm_1}}{2v^2} s^2_{2\theta_\chi}\,, \nonumber \\
\lambda^h_{12} & = \frac{-\lambda_3 + \lambda^\chi_2}{2} s_{2\theta_\chi} -\frac{m^2_{H^\pm_2} - m^2_{H^\pm_1}}{2v^2} c_{2\theta_\chi} s_{2\theta_\chi}\,, \nonumber \\ 
  \lambda_{L} & = \lambda_3 + \lambda_4 + \lambda_5\,. \label{eq:Higgs_triple}
\end{align}
Here, according to Eq.~(\ref{eq:theta_chi}),  we have used the mixing angle $\theta_\chi$ instead of $\mu_\chi$.

  \subsection{Gauge couplings to the $Z_2$-odd particles}
  
  To study the $t\to q (\gamma, Z)$ and $\ell\to \ell' \gamma$ processes, which arise from $\gamma$- and $Z$-penguin diagrams mediated by $H^\pm_{1,2}$ and $S_I(A_I)$ in the loops, we need to know the gauge couplings to these scalars. The kinetic terms of $H_I$ and $\chi^\pm$ in the $SU(2)_L\times U(1)_Y$ gauge symmetry are written by
  \begin{equation}
  {\cal L}_{\rm  kin} \supset (D_\mu H_I)^\dagger D^\mu H_I + (D_\mu \chi^+)^\dagger D^\mu \chi^+\,,
  \end{equation}
  where the covariant derivatives of the scalar fields are given by:
  \begin{align}
  D_\mu H_I & = \left( \partial_\mu + i \frac{g}{2} \vec{\tau}\cdot \vec{W}_\mu + i \frac{g'}{2} B_\mu \right) H_I
  \,, \nonumber \\
  D_\mu\chi^+ & = \left( \partial_\mu + i g' B_\mu \right) \chi^+
  \,. 
  \end{align}
 If we parametrize  the photon and $Z$-gauge boson states as
  \begin{align}
  A_\mu &=  c_W B_\mu + s_W W^3_\mu \,, \nonumber \\
  Z_\mu &= -s_W B_\mu + c_W W^3_\mu\,,
  \end{align}
  with $c_W(s_W)=\cos\theta_W(\sin\theta_W)$ and $\theta_W$ being the Weinberg's angle, the gauge couplings to $H^+_i$, $S_I$, and $A_I$ are obtained  as:
  \begin{align}
  {\cal L}_{\gamma,Z,W^\pm}  
  \supset &~ 
  i e A^\mu \sum^2_{i=1} (\partial_\mu H^-_i H^+_i - H^-_i \partial_\mu H^+_i) \nonumber \\
  &+   i \frac{g}{2c_W} c^Z_{ij} Z^\mu   \left( \partial_\mu H^-_i H^+_j - H^-_i \partial_\mu H^+_j \right)  - \frac{g}{2c_W} Z^\mu ( \partial_\mu A_I H_I - A_I \partial_\mu H_I) \nonumber \\
  &+ i \frac{g \xi_i}{\sqrt{2}} W^{+\mu} \left[  \left(\partial_\mu H^-_i (S_I + i A_I) -H^-_i \partial_\mu(S+iA_I)  \right)
  + \mbox{H.c.} \right]\,, \label{eq:gauge}
  \end{align}
where $\xi_1=c_{\theta_\chi}$, $\xi_2=-s_{\theta_\chi}$, and the coefficients $c^Z_{ij}$ are given by
 \begin{equation}
 c^Z_{11}  = c^2_{\theta_\chi}-2 s^2_W\,,~c^Z_{12} =c^Z_{21}= c_{\theta_\chi} s_{\theta_\chi}\,,~ c^Z_{22} = s^2_{\theta_\chi} -2 s^2_W\,.
 \end{equation}
 
 In addition to the emission from $H^\pm_i$ and $S_I(A_I)$, the photon and $Z$-gauge boson can be emitted from the $B$ quark  in the loop diagrams. If we write the covariant derivative of $B$ quark  to be $D_\mu B= (\partial_\mu + i g' Q_B B_\mu) B$, the  gauge couplings of the $B$ quark are given by:
 \begin{equation}
 {\cal L}_{VBB} = - eQ_B \overline B \gamma_\mu B A^\mu + \frac{g Q_B s^2_W }{c_W} \overline B \gamma_\mu B Z^\mu\,.
 \end{equation}
Since $B$ is a $SU(2)_L$ singlet,  there is no  charged-current interaction with the $W$-gauge boson. 
  
\subsection{Yukawa couplings}

In addition to the Higgs and gauge couplings, the Yukawa interactions of the SM fermions to $Z_2$-odd particles are also important for the rare FCNC processes.  According to the representations and charge assignments of $Z_2$-odd particles, the relevant Yukawa interactions and mass term are:
\begin{align}
- {\cal L}_{Y} 
=&~
\overline L Y^{\ell'} H \ell'_R + \overline{ N_{kL}} {\bf y}^\ell_{1k}  \ell'_R \chi^+
+ \overline L {\bf y}^\ell_{2 k} \widetilde{H_I} N_{k R}  +  \overline{B_L} {\bf y}^{B}_1 u_R \chi^- 
\nonumber \\
&+ 
\overline{Q_L} {\bf y}^B_2 H_I B_R + m_{B} \overline B_L \bar B_R + m_{N_k} \overline{N_{kL}} N_{kR} + \mbox{H.c.}
\,,  \label{eq:L_Yu}
\end{align}
where we have suppressed the flavor indices,  ${\bf y}^\ell_{j k}$ and ${\bf y}^{B}_{j}$ ($j=1,2$) carry the lepton and quark flavors, $L^T=(\nu_{\ell'}, \ell')_L$ and $Q^T=(u, d)_L$ are the lepton and quark doublets in the SM, respectively, and $\widetilde{H_I} \equiv i \tau_2 H^*_I$.  In terms of the physical states, the Yukawa couplings of fermions to $H^\pm_i$ and $S_I(A_I)$  can be written as:
\begin{align}
-{\cal L}_Y 
\supset &~ 
\overline u \left( C^i_{BL} P_L + C^i_{BR} P_R \right) B H^+_i 
+ \overline \ell \left( C^i_{N_k L} P_L + C^i_{N_{k} R} P_R  \right)N_k H^{-}_{i}  
\nonumber \\
 & + \frac{1}{\sqrt{2}}\overline d  V^\dagger_{\rm CKM} {\bf y}^B_{2} P_R B (S_I + i A_I) 
 + \frac{1}{\sqrt{2}} \overline \nu {\bf y}^\ell_{2k} P_R N_k (S_I - i A_I)
 + \mbox{H.c.}
 \,, \label{eq:Yu}
\end{align}
where the weak states of up-type quarks are chosen to align to their physical states,  $V_{\rm CKM}$ is the Cabibbo-Kobayashi-Maskawa (CKM) matrix,  and the couplings $C^i_{BL,BR}$ and $C^i_{N_k L, N_k R}$ are given by:
 \begin{align}
 C^1_{BL} & = {\bf y}^B_1 s_{\theta_\chi} \,, ~C^1_{BR} ={\bf y}^B_2 c_{\theta_\chi}\,, ~ C^2_{BL} = {\bf y}^B_{1} c_{\theta_\chi}\,, ~C^2_{BR} = - {\bf y}^{B}_{2} s_{\theta_\chi} \,,\nonumber \\
  C^1_{N_kL} & = {\bf y}^\ell_{1k} s_{\theta_\chi} \,, ~C^1_{N_k R} =-{\bf y}^\ell_{2k} c_{\theta_\chi}\,, ~ C^2_{N_k L} = {\bf y}^\ell_{1k} c_{\theta_\chi}\,, ~C^2_{N_kR} =  {\bf y}^{\ell}_{2k} s_{\theta_\chi} \,.
 \end{align}

\section{ Phenomenology } \label{sec:phenomenology}

Based on the introduced interactions, we formulate in this section the expressions for the processes of interest, such as $t\to q (h,\gamma, Z)$, $h\to \ell \ell'$, radiative lepton decays, lepton $g-2$, and the oblique parameters, which can be related to the correction of $W$ mass. We note that since the calculation for $t\to q g$ is similar to that for $t\to q \gamma$ and, by neglecting the small different factor, the branching ratio for $t\to q g$ can be approximately estimated as: 
\begin{equation}
BR(t\to q g) \sim  \frac{\alpha_s}{\alpha}  C_F BR(t\to q \gamma) 
\,, \label{eq:tqgluon}
\end{equation}
where $C_F=4/3$, $\alpha=e^2/4\pi$, and $\alpha_s=g^2_s/4\pi$. Using $\alpha_s/\alpha\sim 14.2$, it can be seen that the branching ratio of $t\to q g$ is roughly one order of magnitude larger than that of $t\to q \gamma$. In the following, we just focus on the $t\to q \gamma$ analysis.

\subsection{Top-FCNC processes}

\begin{figure}[phtb]
\begin{center}
\includegraphics[scale=0.8]{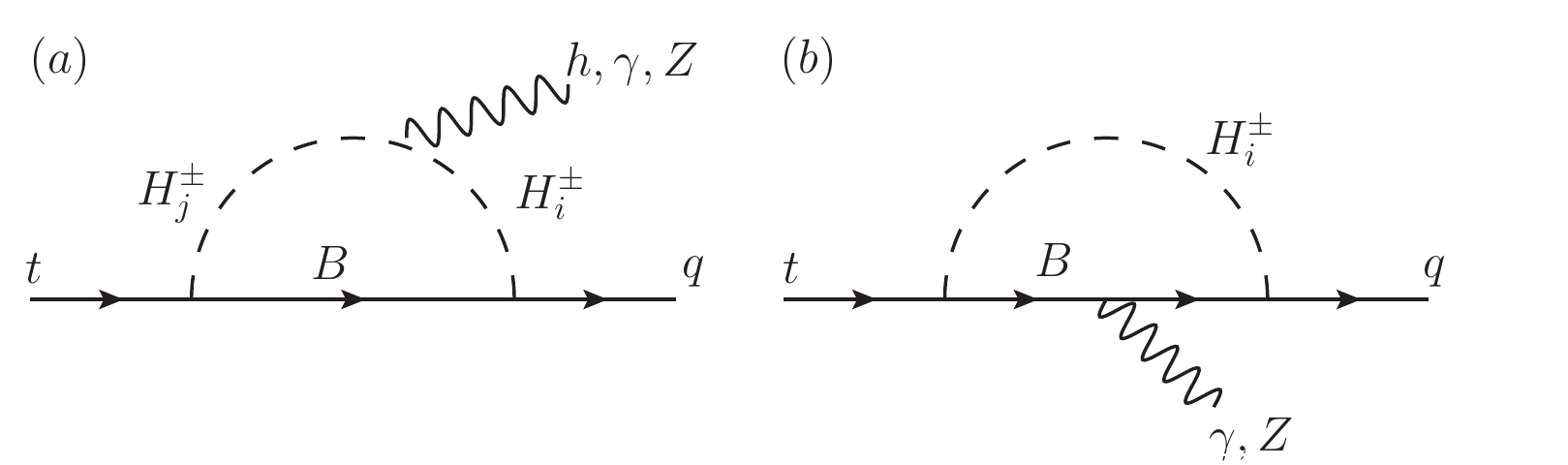}
 \caption{ Feynman diagrams for $t\to q (h,\gamma,Z)$ induced by $H^\pm_i$ and $B$. }
\label{fig:tq_hV}
\end{center}
\end{figure}

In this subsection, we derive the loop-induced effective interactions for top-FCNC processes, where the current upper limits are shown in Table~\ref{tab:top_FCNC}. Because we only introduce a down-type $B$ quark, from the Yukawa interactions in Eq.~(\ref{eq:Yu}), it can be seen that the loop-induced top-FCNCs can only arise from the inert charged Higgses, where the Feynman diagrams are shown in Fig.~\ref{fig:tq_hV}.  Since the photon couplings are Higgs-flavor diagonal, $H^\pm_i = H^\pm_j$ for the $\gamma$-penguin  in Fig.~\ref{fig:tq_hV}(a).  

\begin{table}[htp]
\caption{Experimental upper limits of $t\to q (h, \gamma, Z)$~\cite{PDG2022}. }
\begin{center}
\begin{tabular}{c|cccc}  \hline \hline
 Channel & $t \to uh$ & $t\to ch$ & $t\to q \gamma$ & $t\to q Z$ \\ \hline
  Exp. UL & ~~~$1.2 \times 10^{-3}$~~~ &~~~ $1.1 \times 10^{-3}$~~~ &~~~ $1.8 \times 10^{-4}$~~~& ~~~$5 \times 10^{-4}$ \\ \hline \hline
\end{tabular}
\end{center}
\label{tab:top_FCNC}
\end{table}%
 
In terms of the chiral structures of the initial- and final-state quarks, 
the  effective interactions for $t\to q (h, \gamma, Z)$ can be parametrized as~\cite{Chen:2022dzc}:
 \begin{align}
 {\cal L}_{t\to q (h, \gamma,Z)} 
 =&~ 
 - C^h_{L} \overline q P_L t\,  h - C^h_R \overline q P_R t\, h
 - \frac{ 1 }{m_t}  \overline q i \sigma_{\mu \nu} \epsilon^{\mu*}_\gamma k^\nu  ( B^\gamma_L P_L   +  B^\gamma_R P_R ) t   
 \nonumber \\
& - \bar q \gamma_\mu  \left( A^Z_L P_L + A^Z_R P_R\right ) t Z^\mu 
- \frac{1}{ m_t} \bar q i\sigma_{\mu \nu} \epsilon^{\mu *}_Z k^\nu \left( B^Z_L P_L + B^Z_R P_R   \right) t 
\,, \label{eq:eff}
  \end{align}
  where $\epsilon_V$ denotes the polarization of a vector gauge boson.
 Using the introduced interactions, we can find the relations among the effective coefficients
 and the parameters in the model.  The branching ratios of $t\to q (h,\gamma, Z)$ are then given by: 
 \begin{align}
  Br(t\to q \gamma) 
  =& 
  \frac{ m_t}{ 16 \pi  \Gamma_t }  \left( |B^\gamma_L|^2 + |B^\gamma_R|^2\right)
  \,, \nonumber \\
  Br(t\to q h) 
  =&  
  \frac{m_t}{ 32 \pi  \Gamma_t }  \left(1-\frac{m^2_h}{m^2_t} \right)^2 \left( |C^h_L|^2 + |C^h_R|^2\right)
  \,, \nonumber \\
  Br(t\to q Z) 
  =&
  \frac{1}{\Gamma_t} \frac{G_F m_t^3}{ 4 \sqrt{2} \pi} \frac{c^2_W}{g^2} \left( 1 -\frac{m^2_Z}{m^2_t}\right)^2 
  \sum_{\chi=L,R} \left[ |A^Z_\chi - B^Z_{\chi'} |^2 \left( 1+ \frac{ 2m^2_Z}{m^2_t}\right) \right. 
  \nonumber \\
   &  \left. + \left(| B^Z_{\chi}|^2 +2 Re(A^Z_\chi - B^Z_{\chi'}) B^{Z*}_{\chi'}) \right)\left( 1- \frac{ m^2_Z}{m^2_t}\right)\right]
   \,,
  \end{align}
  where $\Gamma_t$ is the top-quark width, and $\chi'$ has to be chosen to have the opposite chirality to $\chi$, i.e., $B^Z_{\chi'}=B^Z_{R(L)}$ when $A^Z_{\chi}=A^{Z}_{L(R)}$.

 Since we focus on the scenario that $m_B\sim$ TeV and $m_{H_I,\chi}\sim {\cal O}(v)$, for simplicity we neglect small factors, such as $m^2_{H^\pm_i}/m^2_B$ and $m^2_{t,h,Z}/m^2_B$.  As a result, the effective coefficients from Fig.~\ref{fig:tq_hV}(a) and (b) for $t\to q \gamma$ are:
\begin{align}
B^\gamma_R & \approx  \sum^2_{i=1}\frac{ e (C^i_{BR})_q (C^i_{BL})_3 }{(4\pi)^2} \sqrt{r_t}\left( J^1_\gamma(r_{H^\pm_i}, r_t, 0 ) - Q_B J^2_\gamma( r_{H^\pm_i}, r_t, 0)\right)\,, \nonumber \\
B^\gamma_L & \approx \sum^2_{i=1}\frac{e  (C^i_{BL})_q (C^i_{BR})_3}{(4\pi)^2}  \sqrt{r_t} \left( J^1_\gamma(r_{H^\pm_i}, r_t, 0 ) - Q_B J^2_\gamma( r_{H^\pm_i}, r_t, 0)\right)\,;   \label{eq:tqgamma}
\end{align}
those for $t\to q h$ are:
\begin{align}
C^h_R &\approx  \frac{v}{(4\pi)^2 m_B} \sum^2_{i,j=1} (C^j_{BL})_3 \lambda^h_{ji} J_h(r_{H^\pm_i},r_{H^\pm_j},r_t,r_h) (C^i_{BR})_q \,, \nonumber \\
C^h_L &\approx \frac{v}{(4\pi)^2 m_B} \sum^2_{i,j=1} (C^j_{BR})_3 \lambda^h_{ji} J_h(r_{H^\pm_i},r_{H^\pm_j},r_t,r_h) (C^i_{BL})_q \,; \label{eq:tqh}
\end{align}
and those for $t\to q Z$ are:
\begin{align}
B^Z_{R}&\approx  \frac{g  \sqrt{r_t}}{2c_W (4\pi)^2}    \sum^2_{i,j=1} (C^j_{BL})_3   (C^i_{BR})_q \left( c^Z_{ji}  J^1_Z(r_{H^\pm_i},r_{H^\pm_j},r_t,r_Z)- 2 s^2_W  Q_B \delta_{ij}  J^2_\gamma(r_{H^\pm_i}, r_t, r_Z) \right)\,, \nonumber \\
B^Z_{L}&\approx   \frac{g  \sqrt{r_t}}{2c_W (4\pi)^2}    \sum^2_{i,j=1} (C^j_{BR})_3   (C^i_{BL})_q \left( c^Z_{ji}  J^1_Z(r_{H^\pm_i},r_{H^\pm_j},r_t,r_Z)- 2 s^2_W Q_B  \delta_{ij}  J^2_\gamma(r_{H^\pm_i}, r_t, r_Z) \right)\,, \nonumber \\
A^Z_R & \approx  \frac{g s^2_W Q_B}{c_W (4\pi)^2}  \sum^2_{i=1}(C^i_{BL})_3 (C^i_{BL})_q J^2_Z(r_{H^\pm_i}, r_t, r_Z) +B^Z_L\,, \nonumber \\
%
A^Z_L & \approx  \frac{g s^2_W Q_B}{c_W (4\pi)^2}  \sum^2_{i=1}(C^i_{BR})_3 (C^i_{BR})_q J^2_Z(r_{H^\pm_i}, r_t, r_Z) +  B^Z_R\,; \label{eq:tqZ}
%
\end{align}
where $r_f \equiv m^2_f/m^2_B$, and the loop integrals are defined as:
\begin{align}
J^1_\gamma(a,b,c)&=\int^1_0 dx_1 \int^{x_1}_0 dx_2 \frac{1-x_1 }{1 - x_1 +a x_1 - b x_2 (1-x_1)+ c x^2_2}\,,\nonumber\\
J^2_\gamma(a,b,c)&=\int^1_0 dx_1 \int^{x_1}_0 dx_2 \frac{x_1 }{ x_1 +a (1-x_1) - b  x_2 (1-x_1) + c x^2_2}\,, \nonumber \\
J_h(a,b,c,d)&= \int^1_0 dx_1 \int^{x_1}_0 dx_2 \frac{1 }{ 1 -  (1-a) x_1 + (b-a) x_2 - c x_2 (1-x_1) +d x^2_2}\,, \nonumber \\
 J^1_Z(a,b,c,d)&= \int^1_0 dx_1 \int^{x_1}_0 dx_2 \frac{1 -x_1 }{ 1 -  (1-a) x_1 + (b-a) x_2 - c x_2 (1-x_1) +d x^2_2}\,, \nonumber \\
 J^2_Z(a,b,c)&= \int^1_0 dx_1 \int^{x_1}_0 dx_2 \frac{1 }{x_1 +a (1-x_1) - b  x_2 (1-x_1) + c x^2_2}\,.
\end{align}
With the limits $a,b,c,d\ll 1$, the loop integrals can be simplified to be: $J^1_\gamma\approx J^2_\gamma\approx J^1_Z\approx 1/2$ and $J^2_Z\approx 1$.  We note that the $\sqrt{r_t}=m_t/m_B$ factor is retained, where the $m_t$ does not arise from the mass insertion in the top-quark line but is used to fit  the parametrization in Eq.~(\ref{eq:eff}). 

Since the photon couplings to $H^\pm_{i}$ are charged Higgs flavor diagonal, unlike the $Z$ couplings to $H^\pm_{i}$ which involve the $Z$-$H^\pm_1$-$H^\mp_2$ interactions, we have $B^\gamma_{R,L}\approx 0$ when the limit of $m_{H^\pm_i}/m_B \approx 0$ is taken because of a strong cancellation between $H^\pm_1$ and $H^\pm_2$. In addition, although $B^Z_{R,L}$ in Eq.~(\ref{eq:tqZ}) can avoid the cancellation in the limit of $m_{H^\pm_i}/m_B \approx 0$, for $c^Z_{12}$ and $(C^{1,2}_{BR(BL)})_3 (C^{2,1}_{BL(BR)})_q$ all have $\theta_\chi$ dependence and $m_t/(2 m_B)\sim s^2_W Q_B$ for $m_B\sim 1$ TeV, $B^Z_{R,L}$ are smaller than other effective coefficients and lead to subleading effects. Therefore, it is the first term in $A^{Z}_{R(L)}$ that makes the dominant contribution.

\subsection{Lepton flavor-conserving and -violating Higgs decays }

\begin{figure}[phtb]
\begin{center}
\includegraphics[scale=0.8]{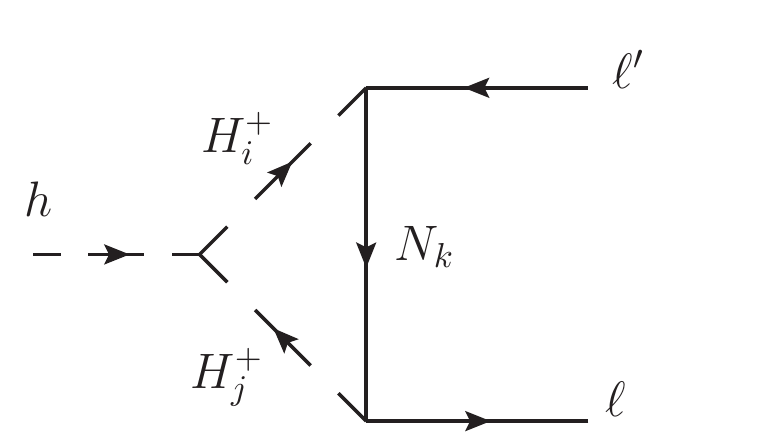}
 \caption{ Feynman diagram for $h\to \ell \overline{\ell'}$ induced by $H^\pm_i$. }
\label{fig:HFV}
\end{center}
\end{figure}

 The $h$-$H^-_i$-$H^+_j$ couplings make important  contributions not only to the $t\to qh$ processes but also to the $h\to \ell \ell'$ processes, where the Feynman diagram for $h\to \ell \ell'$ is sketched in Fig.~\ref{fig:HFV}, and the current upper limits are given in Table~\ref{tab:hellellp}.  The effective $h$-$\ell$-$\ell'$ interactions can be written as:
 \begin{equation}
{\cal L}_{h\ell \ell'} 
= 
- \overline\ell \left( C^R_{\ell \ell'} P_R + C^L_{\ell \ell'} P_L \right) \ell' h\,.
\end{equation}

\begin{table}[htp]
\caption{Experimental upper limits on $h\to \ell \ell'$~\cite{PDG2022}. }
\begin{center}
\begin{tabular}{c|cccc}  \hline \hline
 Channel & $h \to e^+ e^-$ & $h \to e \mu$ &  $h\to e \tau$ & $h\to \mu \tau$ \\ \hline
  Exp. UL & ~~~$3.6 \times 10^{-4}$~~~ &~~~ $6.1 \times 10^{-5}$~~~ &~~~ $2.2 \times 10^{-3}$~~~& ~~~$1.5 \times 10^{-3}$ \\ \hline \hline

\end{tabular}
\end{center}
\label{tab:hellellp}
\end{table}%

Using the Higgs trilinear couplings shown in Eq.~(\ref{eq:Higgs_triple}) and the new Yukawa couplings in Eq.~(\ref{eq:Yu}), the  effective coefficients $C^{R,L}_{\ell \ell'}$  in the model are obtained as:
\begin{align}
C^{R}_{\ell \ell'} & = \frac{1}{(4\pi)^2} \sum^2_{k=1}  \frac{v}{m_{N_k}} \sum^2_{i,j=1}  (C^i_{N_k L})_{\ell'} \lambda^h_{ij} J^{ij}_{hk} (C^j_{N_k R})_\ell  \nonumber \\
&= \frac{1}{(4\pi)^2} \sum^2_{k=1} \frac{v}{m_{N_k}}\left[ \frac{s_{2\theta_\chi}}{2}\left(\lambda^h_{22} J^{22}_{hk} - \lambda^h_{11} J^{11}_{hk}\right)   - c_{2\theta_{\chi}} \lambda^h_{12}  J^{12}_{hk} \right]  ({\bf y}^\ell_{1k})_\ell ({\bf y}^\ell_{2k})_{\ell'} \,,\nonumber \\
%
C^L_{\ell \ell'} & = \frac{1}{(4\pi)^2} \sum^2_{k=1}  \frac{v}{m_{N_k}} \sum^2_{i,j=1}  (C^i_{N_k R})_{\ell'} \lambda^h_{ij} J^{ij}_h (C^j_{N_k L})_\ell  \nonumber \\
&= \frac{1}{(4\pi)^2} \sum^2_{k=1} \frac{v}{m_{N_k}}\left[ \frac{s_{2\theta_\chi}}{2}\left(\lambda^h_{22} J^{22}_{hk} - \lambda^h_{11} J^{11}_{hk}\right)   - c_{2\theta_{\chi}} \lambda^h_{12}  J^{12}_{hk} \right]  ({\bf y}^\ell_{2k})_\ell ({\bf y}^\ell_{1k})_{\ell'} \,,
\end{align}
\begin{equation}
J^{ij}_{hk} = \int^1_0 dx_1 \int^{x_1}_{0} dx_2 \frac{1}{1-(1- r^k_{H^\pm_i}) x_1 + (r^k_{H^\pm_j} -r^k_{H^\pm_i})x_2 + r^k_h x_2 (x_2-x_1)}\,, 
\end{equation}
with $r^k_F \equiv m^2_{F}/m^2_{N_k}$.  It can be seen that $C^{R}_{\ell \ell'}=C^L_{\ell' \ell}$.  The branching ratio of the $h\to \ell \ell'$ decay is then:
\begin{equation}
BR(h\to \ell \ell')
= 
\frac{\zeta_{\ell\ell'} m_h}{ 16 \pi\Gamma_h} 
\left( |C^R_{\ell \ell'}|^2 + |C^L_{\ell \ell'}|^2\right)
\,,
\end{equation}
where $\zeta_{\ell \ell} = 1$ and $\zeta_{\ell \ell'} = 2$. 

\subsection{Radiative lepton flavor violation and lepton $g-2$ }

The radiative LFV processes can be induced via the $\gamma$-penguin diagram  mediated by the inert charged Higgs boson.  Using the Yukawa couplings in Eq.~(\ref{eq:Yu}) and gauge coupling in Eq.~(\ref{eq:gauge}),  the loop-induced effective $\ell$-$\ell'$-$\gamma$ interactions can be written as:
\begin{equation}
{\cal L}^\gamma_{\ell' \ell \gamma}  
=    
e   \, \overline{\ell'} i \sigma_{\mu \nu} 
\left(  C^\gamma_{R \ell' \ell} P_R + C^\gamma_{L \ell' \ell } P_L \right) 
\ell \epsilon^{\mu*}_\gamma k^\nu  
~, \label{eq:L_Cgamma}
\end{equation}
where the  effective coefficients in the model are given by:
 \begin{align}
 C^\gamma_{R\ell' \ell} 
 & = 
 \sum^2_{i,k=1} \frac{(C^i_{N_k R})_{\ell'} (C^i_{N_k L})_\ell} {(4\pi)^2 m_{N_k }} J^1_\gamma\left( \frac{m^2_{H^\pm_i}}{m^2_{N_k}},0,0\right)   
 \nonumber \\
 & = 
 -\frac{s_{2\theta_\chi}}{2 (4\pi)^2}\sum^2_{k=1} \frac{({\bf y}^\ell_{2k})_{\ell'} ({\bf y}^\ell_{1k})_\ell}{m_{N_k}} \Delta J^{1k}_\gamma
 \,, \nonumber 
 \\
 C^\gamma_{L\ell' \ell} 
 & =  
 \sum^2_{i,k=1} \frac{(C^i_{N_k L})_{\ell'} (C^i_{N_k R})_\ell}{(4\pi)^2 m_{N_k } } J^1_\gamma\left( \frac{m^2_{H^\pm_i}}{m^2_{N_k}},0,0\right) ~
 \nonumber \\
 &=  
 -\frac{s_{2\theta_\chi}}{2 (4\pi)^2}\sum^2_{k=1} \frac{({\bf y}^\ell_{1k})_{\ell'} ({\bf y}^\ell_{2k})_\ell}{m_{N_k}} \Delta J^{1k}_\gamma
 \,,
 \label{eq:C_litoljga}
\end{align}
with 
\begin{equation}
\Delta J^{1k}_\gamma 
=  
J^1_\gamma\left(\frac{m^2_{H^\pm_1}}{m^2_{N_k}},0,0\right)
- J^1_\gamma\left( \frac{m^2_{H^\pm_2}}{m^2_{N_k}},0,0\right) 
\,. \label{eq:DJ}
\end{equation}
Because the Yukawa couplings of charged scalars to fermions involve left-handed and right-handed states, the chirality flip occurring in the propagator of the $N_k$ fermion line leads to an enhancement factor of $m_{N_k}$.  Therefore, $C^\gamma_{R\ell' \ell}$ and $C^\gamma_{L\ell' \ell}$ are only suppressed by $1/m_{N_k}$ instead of $1/m^2_{N_k}$.  Moreover, for the photon radiative decays, although $m^2_{H^\pm_i}/ m^2_{N_k} \ll 1$, we cannot take $\Delta J^{1k}_{\gamma}\sim 0$ in Eq.~(\ref{eq:DJ}); otherwise, $C^\gamma_{R\ell' \ell}\sim C^\gamma_{L\ell' \ell}\sim 0$. 
The branching ratio of $\ell\to \ell' \gamma$ can be estimated using:
\begin{equation}
 BR(\ell \to \ell' \gamma) \approx
 \frac{\tau_\ell  \alpha m^3_\ell }{4} \left( \left| C^{\gamma}_{R\ell' \ell} \right|^2 +\left| C^{\gamma}_{L\ell' \ell} \right|^2 \right) ~,
 \label{eq:litoljga}
 \end{equation}
 where $\tau_\ell$ denotes the lifetime of $\ell$ lepton.

From the electromagnetic dipole interactions in Eq.~(\ref{eq:L_Cgamma}), the lepton $g-2$ can be obtained in a straightforward way by taking $\ell'=\ell$ and given by:
\begin{align}
\Delta a_\ell 
& = 
\frac{m_\ell}{8\pi^2} \sum^2_{i,k=1} 
\frac{(C^i_{N_k L})_\ell (C^i_{N_k R})_\ell }{ m_{N_k}} 
J^1_\gamma\left( \frac{m^2_{H^\pm_i}}{m^2_{N_k}},0,0\right) 
\nonumber \\
& = 
- \frac{m_\ell s_{2\theta_\chi}}{16 \pi^2} 
\sum^2_{k=1} \frac{({\bf y}^\ell_{1k})_\ell ({\bf y}^\ell_{2k})_\ell}{m_{N_k}} 
\Delta J^1_\gamma 
\,. \label{eq:lepton_gm2}
\end{align}
In general, the above lepton $g-2$ correction can be positive or negative, depending on the signs of $s_{\theta_\chi}$ and $({\bf y}^\ell_{1k(2k)})_\ell$.

\subsection{Oblique parameters and $W$ boson mass}

In addition to enhancing the rare decays in the SM, the newly introduced particles and couplings can make significant contributions to the vacuum polarization tensors,  parametrized by~\cite{Grimus:2008nb}
 \begin{align}
 \Pi^{\mu\nu}_{VV'} (q)= g^{\mu\nu} A_{VV'}(q^2) + q^\mu q^\nu B_{VV'}(q^2) \,,
 \end{align}
where $VV'$ can be the gauge boson pairs $\gamma\gamma$, $\gamma Z$, $ZZ$, and $WW$.  To show the sensitivity to the new physics effects, we can use the oblique parameters $S$, $T$, and $U$, which are related to $A_{VV'}(q^2)$ and defined as~\cite{Peskin:1990zt,Peskin:1991sw,Maksymyk:1993zm}:
 \begin{align}
 \alpha T & = \frac{A_{WW}(0)}{m^2_W}-\frac{A_{ZZ}(0)}{m^2_Z}\,, \nonumber \\
 \frac{\alpha}{4 s_W c_W} S& 
 = \frac{A_{ZZ}(m^2_Z) -A_{ZZ}(0)}{m^2_Z} - \frac{\partial A_{\gamma\gamma}(q^2)}{\partial q^2} \Big{|}_{q^2=0} + \frac{c^2_W -s^2_W}{s_W c_W }  \frac{\partial A_{\gamma Z}(q^2)}{\partial q^2} \Big{|}_{q^2=0} \,, \nonumber \\
 U+ S & = \frac{4 s^2_W}{\alpha} \left(  \frac{A_{WW}(m^2_W)-A_{WW}(0)}{m^2_W} -  \frac{\partial A_{\gamma\gamma}(q^2)}{\partial q^2} \Big{|}_{q^2=0} + \frac{c_W}{s_W}  \frac{\partial A_{\gamma Z}(q^2)}{\partial q^2} \Big{|}_{q^2=0}  \right)\,.
 \end{align}
Since the $U$ parameter can be taken as the effect of a dimension-8 operator and is normally much smaller than $T$ and $S$ that are due to the effects of dimension-6 operators, we take $U\approx 0$ in the model. 

Because the $Z_2$-odd $B$ quark and $N$ are singlets and do not mix with the SM fermions, they do not contribute to the $T$ and $S$ parameters. Thus, only the inert Higgs doublet $H_I$ and the charged singlet $\chi^\pm$ will affect the oblique parameters.  According to the results obtained in Ref.~\cite{Herrero-Garcia:2017xdu}, the $T$ parameter from $H_I$ and $\chi^\pm$ is given by:
\begin{align}
T
=&~ 
\frac{1}{16  \pi^2 c^2_W s^2_W } \left[ c^2_{\theta_\chi} \left( \theta_{+}(z_{H^\pm_1}, z_{H_I}) +\theta_{+}(z_{H^\pm_1}, z_{A_I})\right) \right.
\nonumber \\
& \left. + s^2_{\theta_\chi} \left( \theta_{+}(z_{H^\pm_2}, z_{H_I}) +\theta_{+}(z_{H^\pm_2}, z_{A_I})\right) - \frac{1}{2} c^2_{\theta_\chi} s^2_{\theta_\chi} \theta_{+}(z_{H^\pm_1}, z_{H^\pm_2}) - \theta_{+}(z_{H_I}, z_{A_I})\right]
\,, \label{eq:T}
\end{align}
with $z_f \equiv m^2_f/m^2_Z$ and
\begin{equation}
\theta_{+}(x, y) = x + y - \frac{2 x y}{x-y} \ln\left(\frac{x}{y} \right) 
\,.
\end{equation}
The $S$ parameter, on the other hand, is:
\begin{align}
S
=&~ 
\frac{1}{\pi} \left[ G(1, z_{S_I},z_{A_I}) + c^2_{\theta_\chi} (1-s^2_{\theta_\chi}) G(1, z_{H^\pm_1}, z_{H^\pm_1}) \right. 
  \nonumber \\
& \left. -s^2_{\theta_\chi} (1+c^2_{\theta_\chi}) G(1, z_{H^\pm_2},z_{H^\pm_2}) + 2 s^2_{\theta_\chi} c^2_{\theta_\chi} G(1, z_{H^\pm_1}, z_{H^\pm_2}) \right]
\,, \label{eq:S}
\end{align}
where 
\begin{align}
G(a,b,c) 
& = 
B_{22}( a,b,c)-B_{22}(0,b,c)
\,, \nonumber \\
B_{22} (a,b,c) 
& = 
- \frac{1}{2} \int^1_0 dx (b x + c (1-x) - a x(1-x))\ln\left(b x + c (1-x) - a x(1-x)\right)
\,.
\end{align}
From the results, it can be seen that when $\chi^\pm$ decouples from  $H^\pm_I$, i.e. $s_{\theta_\chi}=0$, only the inert Higgs doublet contributes to the oblique parameters. 
 
It is known that the relation of $m_W$ to the oblique parameters can be expressed as~\cite{Maksymyk:1993zm}:
 \begin{align}
 m_W  
 & =
 m^{\rm SM}_W \left[ 1 +  \frac{\alpha}{c^2_W-s^2_W} \left( c^2_W T - \frac{S}{2} + \frac{c^2_W-s^2_W}{4s^2_W} U\right)\right]^{1/2} 
 \nonumber \\
 & \simeq   
 m^{\rm SM}_{W} \left[1 + \frac{\alpha}{4 (c^2_W-s^2_W)} \left( 2 c^2_W T -S\right)\right]
 \,, \label{eq:mW}
 \end{align}
where we have neglected $U$ parameter in the linear approximation in the second line as it is much smaller than $T$ and $S$. Taking $s^2_W = 0.2316$, $\alpha=1/129$, and $m^{\rm SM}_W=80.3496$~\cite{deBlas:2022hdk} as the SM inputs, we find that when $T\sim 0.16$ and $S\sim -0.02$,  the $W$ boson mass can be increased to $m_W\approx 80.4267$~GeV. In the section of numerical analysis, we will show the correlation of the parameters with the necessary values of $T$ and $S$ to explain the $m_W$ anomaly. 

\section{Constraints} \label{sec:constraints}

We discuss potentially strict constraints on the model in this section. In addition to  $\mu\to e \gamma$ mentioned earlier, we discuss the severe limits from $\Delta B=2$, $\Delta D=2$, and $h\to \gamma \gamma$ processes. Since the considered mass scale of the inert scalars is set at the EWSB scale, the lightest neutral component, i.e. $S_I$ or $A_I$, can be a DM candidate.  In this case, the Higgs trilinear coupling $h$-$S_I$-$S_I$ or $h$-$A_I$-$A_I$ will be constrained by the DM direct detection through Higgs portal. Nevertheless, because the involved coupling is $\lambda_L$ and it is not directly related to the processes studied in this work, one can take the $\lambda_L\sim 0$ limit to satisfy the nonobservation of DM direct detection.

\subsection{$\Delta B=2$ and $\Delta D=2$ processes}

From Eq.~(\ref{eq:Yu}), it is seen that the down-type quarks couple to $Z_2$-odd $B$ quark and inert neutral scalars.  Thus, $\Delta B=2$ processes induced via box diagrams can give stringent constraints on the Yukawa couplings ${\bf y}^{B}_2$. To estimate the $B_{q'}-\overline B_{q'}$ mixing parameter, we simply use the hadronic effect $\langle \bar B| (\overline{q'} \gamma_\mu P_L b)^2 | B_{q'}\rangle \sim f^2_{B_{q'}} m_{B_{q'}}/3$.  The mass difference between the heavy and light $B$ mesons can be simplified to be:
\begin{align}
\Delta m_{B_{q'}}\sim     \frac{f^2_{B_{q'}} m_{B_{q'}}}{48 \pi^2  m^2_B} \left((V^\dagger_{\rm CKM} {\bf y}^{B}_{2})_{q'} ({\bf y}_2^{B\dagger} V_{\rm CKM})_3 \right)^2  \, ,
\end{align}
where the approximation  $m^2_{S_I, A_I}/m^2_B \sim 0$ appearing in the loop integrals is applied. Due to the fact that $(V_{\rm CKM})_{k3}\ll 1$ ($k=1,2$), we have $({\bf y}_2^{B\dagger} V_{\rm CKM})_3  \sim y^B_{23}$, 
  \begin{align}
  (V^\dagger_{\rm CKM} {\bf y}^{B}_{2})_{d}  & \sim -\lambda y^{B}_{22} + y^{B}_{21} \,, \nonumber \\
   (V^\dagger_{\rm CKM} {\bf y}^{B}_{2})_{s} & \sim y^{B}_{22} + \lambda y^{B}_{21} \,,
  \end{align}
with $\lambda\sim 0.22$ being the Wolfenstein parameter.  If we assume $y^{B}_{21}\sim \lambda y^{B}_{22}$ to avoid the $\Delta m_{B_{d}}$ constraint, $(V^\dagger_{\rm CKM} {\bf y}^{B}_{2})_{s}  \sim y^{B}_{22}$ is then bounded by $\Delta m_{B_s}$. Taking  $f_{B_s}=0.231$~GeV~\cite{Lenz:2010gu} and requiring $\Delta m_{B_s} < \Delta m^{\rm exp}_{B_s} = 1.17 \times 10^{-11}$~GeV~\cite{PDG2022}, the upper limit on $f^{B}_{22} f^{B}_{23}$ can be obtained as:
  \begin{equation}
   \frac{|f^{B}_{22} f^{B}_{23}|}{m_B} <  1.4 \times 10^{-4}
   \,. \label{eq:upper_Bq}
  \end{equation}
If we just consider the contributions from $f^B_{22, 23}$ and ignore the other parameters, the branching ratios of top-FCNCs will be smaller than $10^{-6}$ and lower than the sensitivities of HL-LHC.  Hence, if $f^{B}_{21}$ and $f^{B}_{22}$ are bounded by $\Delta m_{B_d}$ and $\Delta m_{B_s}$, respectively, their effects can be neglected.  

From Eq.~(\ref{eq:Yu}),  it can be found that the couplings to the $Z_2$-odd $B$ quark and inert charged scalars only involve the left-handed up-type quarks. As a result, $f^B_{11}$ and $f^{B}_{12}$ contribute to the $D$-$\overline D$ mixing, and the resulting mixing parameter of $D$ meson can be written as: 
\begin{align}
\Delta m_{D}\sim    - \frac{f^2_{D} m_D}{12 (4 \pi)^2  m^2_B}  \left(y^{B}_{11} y_{12}^{B}  \right)^2  \,,
\end{align}
where we have applied the approximation $m^2_{H^\pm_i}/m^2_B\sim 0$.  With $f_{D}=0.213$~GeV~\cite{PDG2022} and $\Delta m^{\rm exp}_D=6.1 \times 10^{-15}$~GeV, the upper limit on $f^B_{11} f^B_{12}$ is
 \begin{equation}
 \frac{|f^B_{11} f^{B}_{12}|}{m_B} < 1.2 \times 10^{-5}\,. \label{eq:upper_D}
 \end{equation}
Combing the constraints shown in Eqs.~(\ref{eq:upper_Bq}) and (\ref{eq:upper_D}), we conclude that $t\to u(h,V)$ and $t\to c(h,V)$ cannot be simultaneously enhanced up to the sensitivities of HL-LHC. Thus, in the following analysis, we assume $f^B_{11}\ll f^B_{12,13}$ and $f^{B}_{21,22}\ll f^{B}_{23}$ and focus on the $t\to c$ FCNC processes.

\subsection{$h\to \gamma \gamma$}

As stated earlier, the Higgs  trilinear couplings to  $H^\pm_i$ make significant contributions to $t\to q h$ and $h\to \ell \ell'$; it is found that the same couplings can also modify the Higgs to diphoton decay rate.  Since the measurement of $h\to \gamma\gamma$ is approaching a precision level, it is expected that the associated free parameters may suffer from a strict bound. To study the new physics effects on $h\to \gamma\gamma$, let's consider the signal strength of $pp\to h \to \gamma \gamma$ defined by:
 \begin{align}
 \mu_{\gamma\gamma} = \frac{\sigma(pp\to h)}{\sigma(pp\to h)^{\rm SM}} \frac{BR(h\to \gamma \gamma)} {BR(h\to \gamma\gamma)^{\rm SM}} \,,
 \end{align}
where the current world average is $\mu_{\gamma\gamma}=1.10\pm 0.07$~\cite{PDG2022}.

\begin{figure}[phtb]
\begin{center}
\includegraphics[scale=0.8]{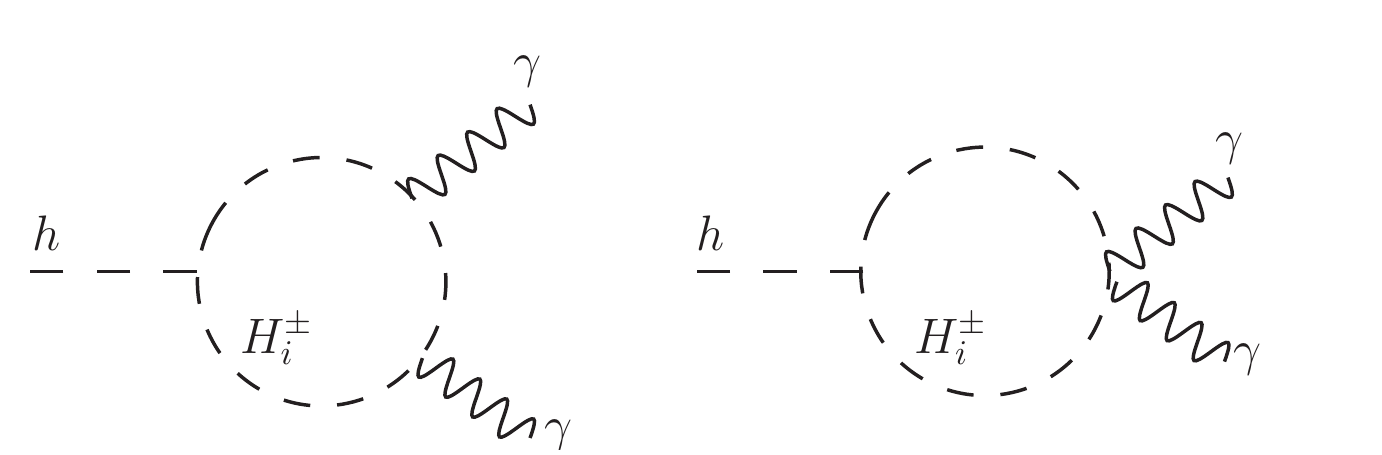}
 \caption{ Feynman diagrams for $h\to \gamma\gamma$ induced by $H^\pm_I$. }
\label{fig:hgaga}
\end{center}
\end{figure}

The loop-induced effective interaction for $h\gamma\gamma$  can be parameterized as:
\begin{equation}
{\cal L}_{hVV}  = \frac{\alpha}{4\pi} \frac{a_{\gamma\gamma}}{m_h}  h F_{\mu \nu} F^{\mu \nu} 
\,,
\end{equation} 
where the Feynman diagrams mediated by $H^\pm_i$ are shown in Fig.~\ref{fig:hgaga}.  The resulting $a_{\gamma\gamma}$ in the model is obtained as:
\begin{equation}
a_{\gamma \gamma}  = \frac{g m_h }{2m_W} \left( a^{\rm SM}_{2\gamma} + \sum^2_{i=1} \frac{\lambda^h_{ii} v m_W}{g m^2_{H^\pm_i}} F_0(\tau_{H^\pm_i})  \right)
\,, 
\end{equation}
where $a^{\rm SM}_{2\gamma}\approx 6.51 - i 0.02$ is the SM, and the function $F_{0}$ is 
 \begin{equation}
 F_0 (\tau)  =  \tau (1 - \tau f(\tau) )\,, 
 \end{equation}
with $\tau \equiv 4 m^2_f/m^2_{H^\pm_i}$ and $f(\tau)= (\arcsin(1/\sqrt{\tau}))^2$.

Because the SM Higgs does not couple to the $B$ quark,  the signal strength for $pp\to h \to \gamma\gamma$ can be simplified as
 \begin{equation}
 \mu_{\gamma\gamma}  \approx \frac{BR(h\to \gamma \gamma)} {BR(h\to \gamma\gamma)^{\rm SM}} \approx \left| 1 + \sum^2_{i=1} \frac{\lambda^h_{ii} v m_W}{ g m^2_{H^\pm_i} a^{\rm SM}_{2\gamma} } F_{0} (\tau_{H^\pm_i}) \right|^2\,, \label{eq:mu2gamma_ICH}
 \end{equation}
where the new physics contribution to the Higgs width $\Gamma_h$ is assumed to be small and neglected in the calculation of $\mu_{\gamma\gamma}$. To suppress the invisible Higgs decay $h\to S_I S_I, A_I A_I$ and to have  $\Gamma_h \approx \Gamma^{\rm SM}_h\approx 4.1$~MeV, we simply take $m_h < 2 m_{S_I, A_I}$ in the model. Using $m_{H^\pm_{1,2}}=(120, 320)$~GeV, the $H^\pm_i$ effect on $\mu_{\gamma\gamma}$ can be estimated to be $-0.125 \lambda^h_{11} - 0.015\lambda^h_{22}$.  Because $\mu_{\gamma\gamma}$ arising from $H^\pm_i$ is proportional to $1/m^2_{H^\pm_i}$, it is seen that the influence of heavy $H^\pm_2$ is small. If we take the allowed range of $\mu_{\gamma\gamma}$ to be $0.9 < \mu_{\gamma\gamma}< 1.2$ and drop the $H^\pm_2$ contribution, then $\lambda^h_{11}$ can be restricted to fall in the range $-0.8<\lambda^h_{11} < 0.5$.  Hence, the $\lambda^h_{11}$ parameter can be bounded by the $h\to \gamma\gamma$ measurement, whereas the region of $\lambda^h_{22}$ is wide.  

\subsection{$\mu\to e \gamma$ decay}

The current experimental upper limits on $\ell\to \ell' \gamma$ are listed in Table~\ref{tab:LFV}~\cite{PDG2022}.  It can be seen that the constraint from $\mu\to e\gamma$ should be much stronger than that from the $\tau$ decays. 

\begin{table}[htp]
\caption{Experimental upper limits on $\ell \to \ell' \gamma$ decays~\cite{PDG2022}. }
\begin{center}
\begin{tabular}{c|ccc}  
 \hline \hline
 Channel & $\mu \to e \gamma$ & $\tau \to e \gamma$ &  $\tau\to \mu \gamma$ 
 \\ 
 \hline
 Exp. UL & ~~~$4.2 \times 10^{-13}$~~~ &~~~ $3.3 \times 10^{-8}$~~~ &~~~ $4.2 \times 10^{-8}$ 
 \\ 
 \hline \hline
\end{tabular}
\end{center}
\label{tab:LFV}
\end{table}%

In order to satisfy the upper limit of $\mu\to e \gamma$,  we assume that the free parameters can satisfy the conditions $C^{\gamma}_{Re \mu}\approx 0$ and $C^{\gamma}_{Le \mu}\approx 0$, where $C^{\gamma}_{R\ell' \ell}$ and $C^{\gamma}_{L\ell' \ell}$ are defined in Eq.~(\ref{eq:C_litoljga}).  As a result, the relevant Yukawa couplings can be related as:
 \begin{align}
 (y^{\ell}_{22})_{e} 
 & \approx 
 - \frac{m_{N_2}}{m_{N_1}} \frac{\Delta J^{11}_{\gamma}}{\Delta J^{12}_\gamma} \frac{(y^{\ell}_{11})_\mu }{(y^\ell_{12})_\mu} 
 \,, \nonumber \\
 (y^{\ell}_{12})_{e} & \approx - \frac{m_{N_2}}{m_{N_1}} \frac{\Delta J^{11}_{\gamma}}{\Delta J^{12}_\gamma} \frac{(y^{\ell}_{21})_\mu }{(y^\ell_{22})_\mu}
 \,.
 \end{align}
If we further take $m_{N_1}= m_{N_2}$, in addition to $BR(\mu\to e\gamma) \approx 0$, we will also have $BR(h\to e \mu) \approx 0$. 

\section{Numerical Analysis} \label{sec:Num}

\subsection{Parameter setting} 

The relevant free parameters introduced in the model are the Yukawa couplings ${\bf y}^B_{1,2}$ and ${\bf y}^{\ell}_{1k, 2k}$, the mixing angle $\theta_\chi$, the masses of inert scalars $m_{H^\pm_i, S_I, A_I}$, the masses of new fermions $m_{B,N_k}$, and the scalar couplings from the scalar potential $\lambda^h_{ij}$ defined in Eq.~(\ref{eq:Higgs_triple}). Although  some constraints on the free parameters have been studied earlier, in the following we make some more considerations to further restrict their ranges before scanning the observables.  
 
\begin{figure}[phtb]
\begin{center}
\includegraphics[scale=0.5]{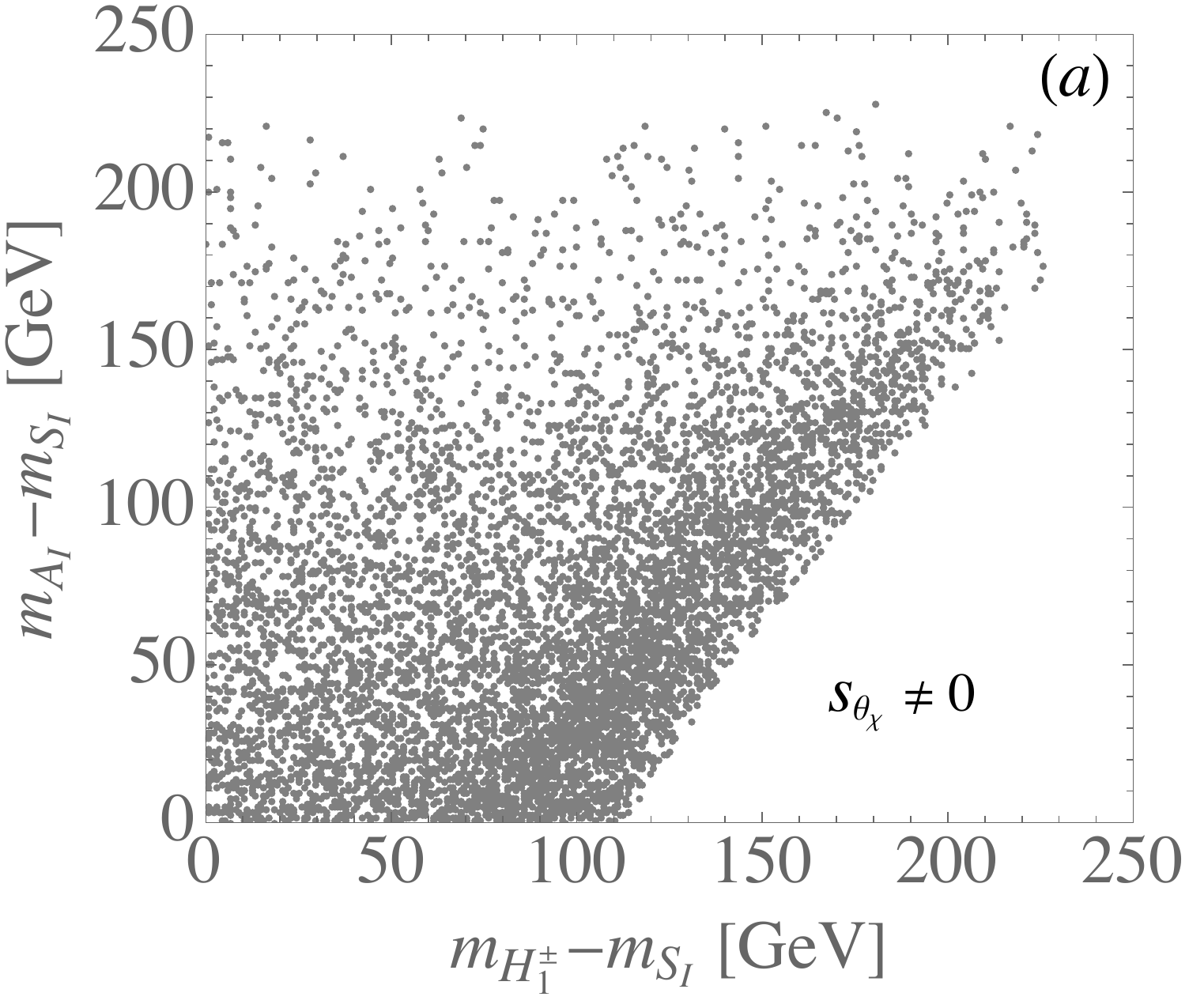}
\includegraphics[scale=0.5]{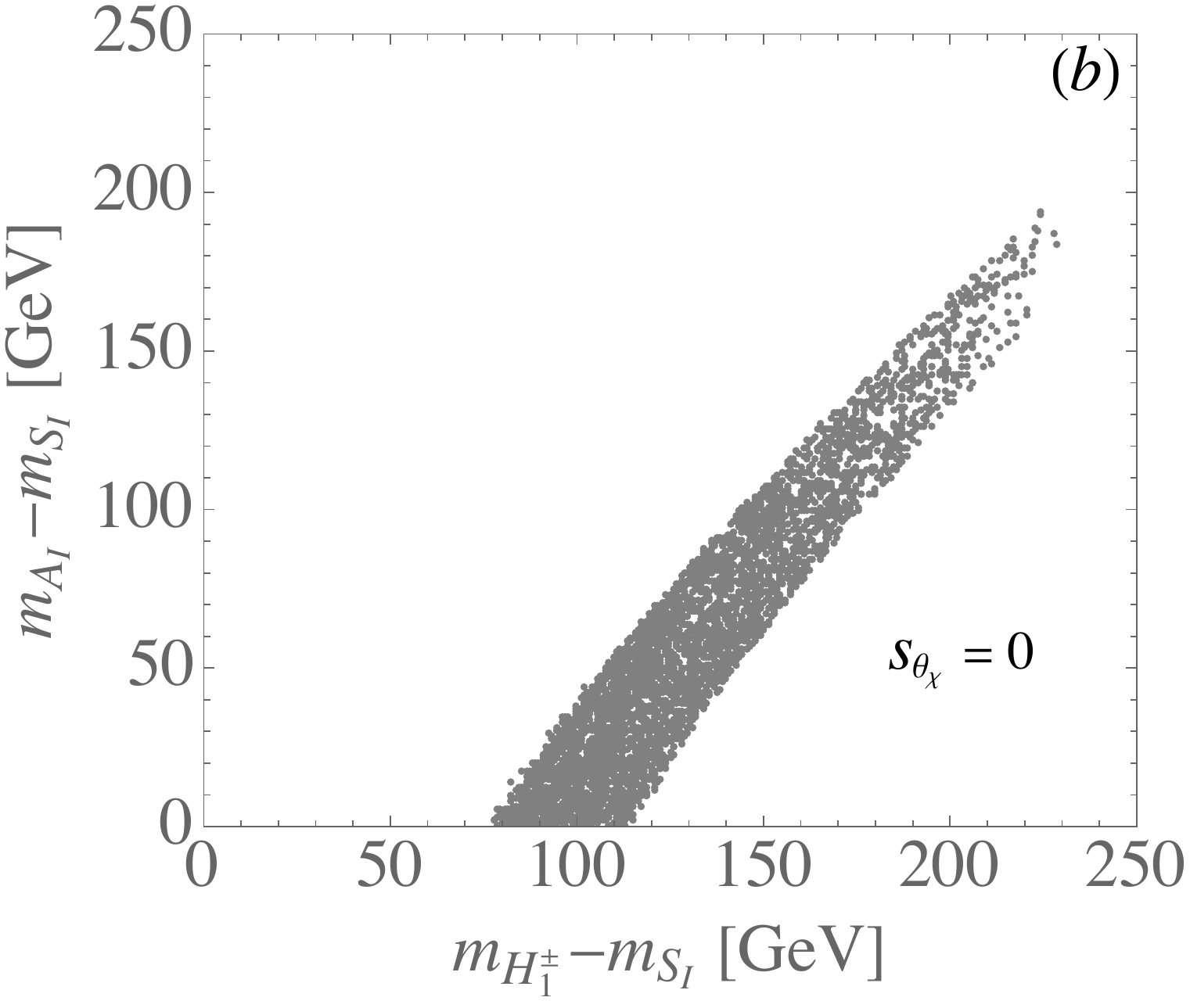}
\includegraphics[scale=0.5]{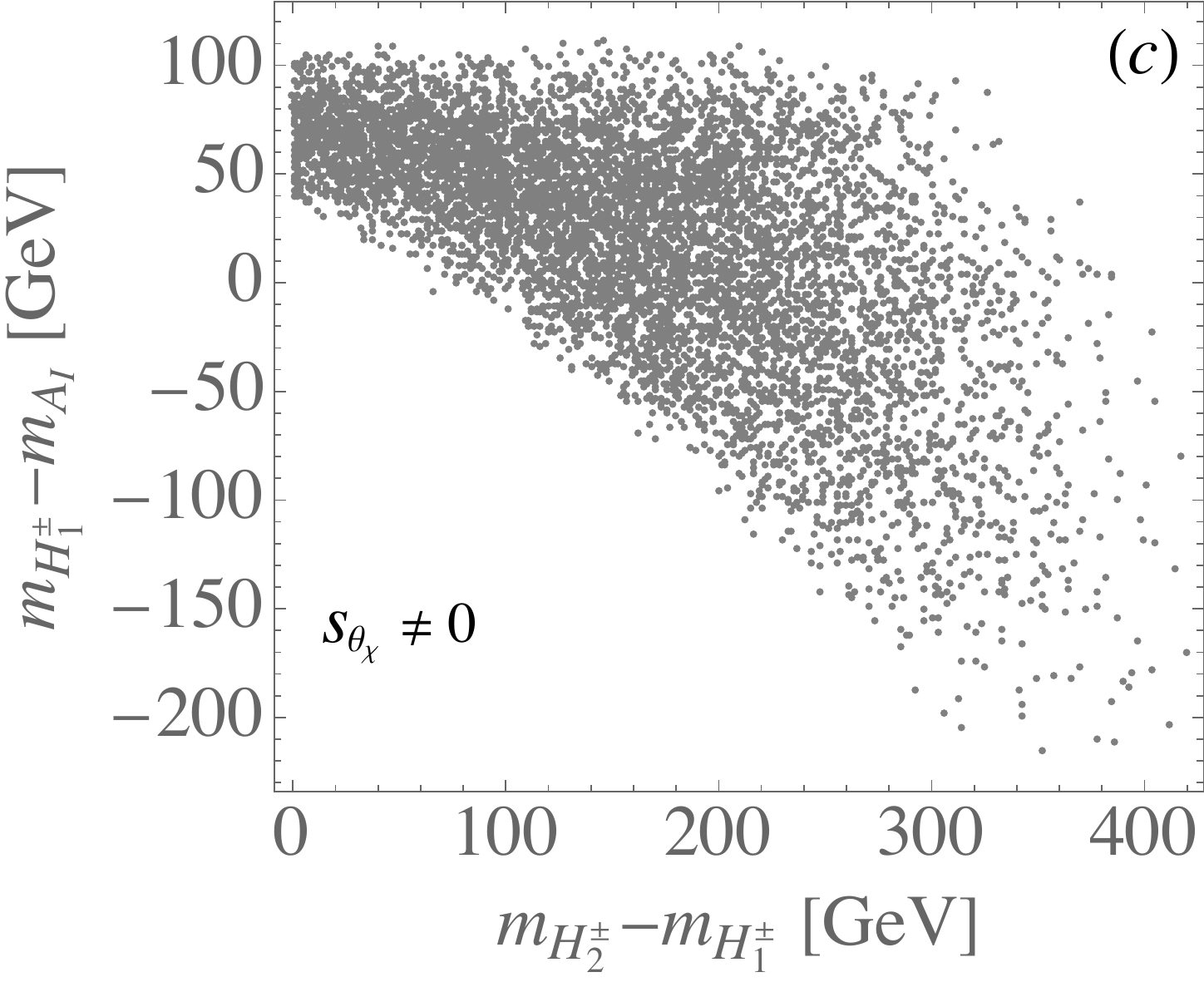}
\includegraphics[scale=0.5]{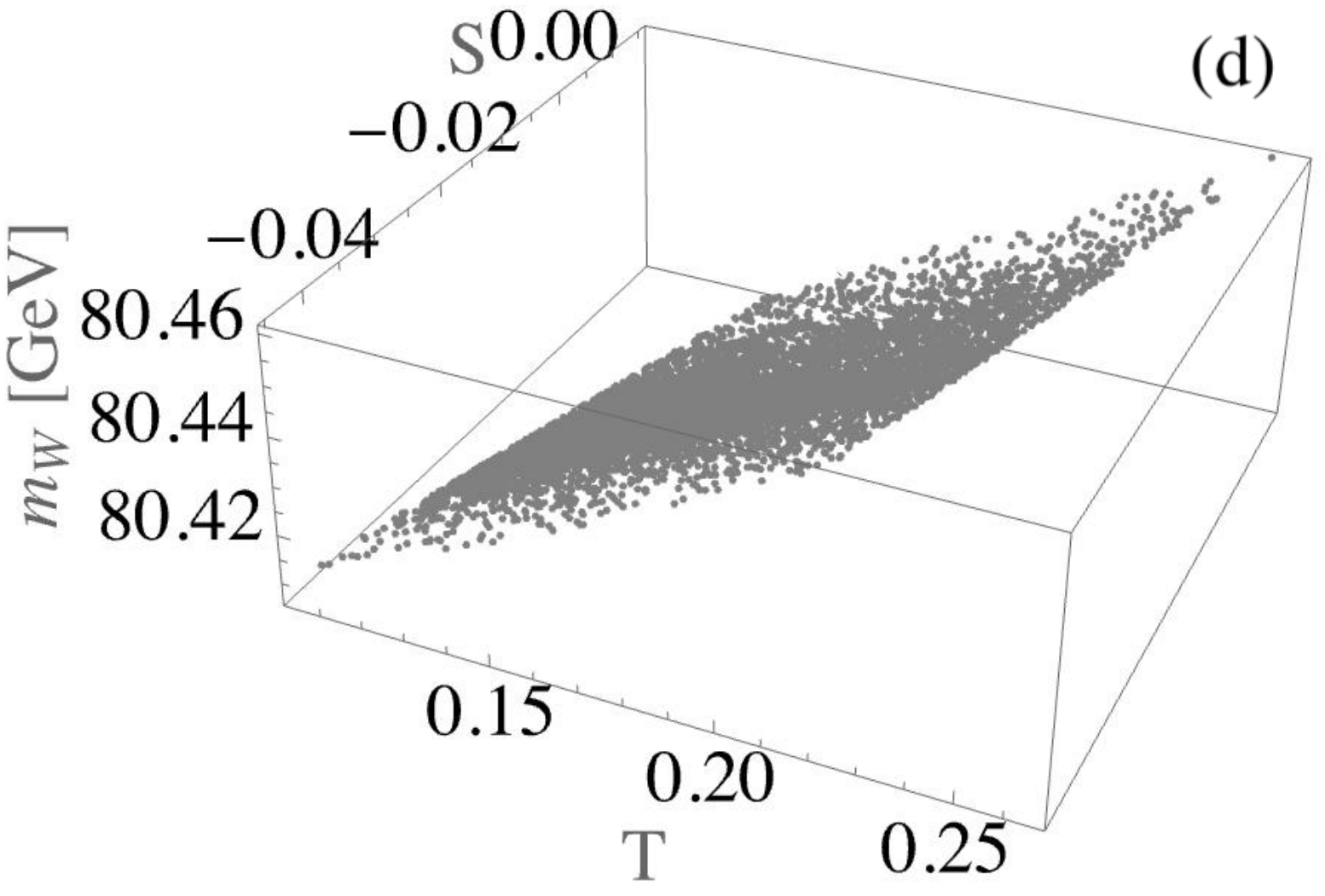}
 \caption{ Correlation between $m_{H^\pm_1}-m_{S_I}$ and $m_{A_I}-m_{S_I}$ for (a) $s_{\theta_\chi}=0$ and (b) $s_{\theta_\chi}\neq 0$;  (c) correlation between $m_{H^\pm_2}-m_{H^\pm_1}$ and $m_{H^\pm_1}-m_{A_I}$, and (d) the resulting $m_W$ as a function of $T$ and $S$ parameters, where $T=0.202\pm 0.056$ and $S=0.100\pm 0.073$ in their $2\sigma$ ranges. }
\label{fig:TS}
\end{center}
\end{figure}

Generally, each of $N$ and $S_I(A_I)$ can be a DM candidate.  Since we concentrate on the scenario where $m_{S_I(A_I)}\sim {\cal O}(m_W)$, as it can fit the observed DM relic density~\cite{Barbieri:2006dq}, we take $m_{N}> m_{S_I(A_I)}$ in this work. To be specific, we assume $m_{S_I} < m_{A_I, H^\pm_i}$ so that $S_I$ is the DM candidate.  Because the oblique parameters are sensitive to the mass differences among  $H^\pm_i$, $S_I$, and  $A_I$, we first discuss the constraint from the $T$ and $S$ parameters. Using Eqs.~(\ref{eq:T}) and (\ref{eq:S}),  we make the scatter plot for the correlation between $m_{H^\pm_1}-m_{S_I}$ and $m_{A_I}-m_{S_I}$ in Fig.~\ref{fig:TS}(a), where we take $s_{\theta_\chi}\in (-1/\sqrt{2}, 1/\sqrt{2})$, $m_{S_I,A_I,H^\pm_1}\in(70, 300)$~GeV, $m_{H^\pm_2}\in (70, 500)$~GeV, and $T=0.202\pm 0.056$ and $S=0.100\pm 0.073$ ~\cite{deBlas:2022hdk} with $2\sigma$ errors as the allowed ranges.  For the purpose of comparison, we show the result with $s_{\theta_\chi}=0$  in Fig.~\ref{fig:TS}(b).   It can be seen that to fit the results of $T$ and $S$, the allowed range of  $m_{H^\pm_1}-m_{S_I}$ with $s_{\theta_\chi}\neq 0$ is broader than that with vanishing $s_{\theta_\chi}$.  The correlation between $m_{H^\pm_2}-m_{H^\pm_1}$ and $m_{H^\pm_1}-m_{A_I}$ is shown in Fig.~\ref{fig:TS}(c). The dependence of $m_W$ on $S$ and $T$ in the model is exhibited in Fig.~\ref{fig:TS}(d).    From the results, we see that the $W$-mass anomaly can be explained if $-0.05<S<0.02$ and $0.10< T<0.26$. 

In addition to the constraints from $\Delta B=2$ and $\Delta D=2$, we can also use the perturbative unitarity constraint to bound the Yukawa couplings, where the upper limits  are required to be $|{\bf y}^{B}_{1,2}|, |{\bf y}^\ell_{1k,2k}| < \sqrt{4\pi}$~\cite{Castillo:2013uda}.  For the mass upper limit  of the $Z_2$-odd quark,  we can apply the constraints on the stop and sbottom with the $R$-parity conserving supersymmetry (SUSY).  Using  the data with an integrated luminosity of 139 fb$^{-1}$ at $\sqrt{s}=13$ TeV~\cite{ATLAS:2021hza}, the mass below $1$~TeV has been excluded by ATLAS when the neutralino mass is around 100~GeV. Therefore, we impose $m_B > 1$~TeV.  Although there is no strict limit on the neutral singlet lepton, we take $m_{N} > 1$~TeV in the numerical analysis. 

To simplify the parameter scan, the ranges of parameters satisfying the theoretical requirements and the experimental bounds are chosen as follows:
 \begin{align}
& y^B_{12, 13, 23} \in (-3.5,\, 3.5)\,,~  {\bf y}^{\ell}_{1k,2k} \in (-0.5,\, 0.5)\,, ~ \lambda_3\in (1,\, 5)\,,~ s_{\theta_\chi} \in \left(-\frac{1}{\sqrt{2}},\, \frac{1}{\sqrt{2}} \right)\,, \nonumber  \\
 &m_{B}\in (1, 1.5)~\text{TeV}\,, ~m_{N_1}\in (1, 2)~\text{TeV} \,,~ m_{N_2}=m_{N_1} + 300~ \text {GeV}\,,
 \label{eq:limits}
 \end{align}
where we take  $y^{B}_{11,21,22}\approx 0$ in order not to upset the $B_{q'}-\bar{B}_{q'}$  and $D-\bar{D}$ mixing phenomena, as discussed before.  To satisfy the constraint from $h\to \gamma\gamma$, we set $\lambda^h_{11}\approx 0$.  As a result, we have:
 \begin{equation}
 \lambda^\chi_2 = s_{\theta_\chi} c_{\theta_\chi} \left( \frac{m^2_{H^\pm_2}-m^2_{H^\pm_1}}{v^2} - \lambda_3 \frac{c^2_{\theta_\chi}}{s^2_{\theta_\chi}}\right)
 \,. \label{eq:lambchi2}
 \end{equation} 
Thus, in the numerical calculations, we use Eq.~(\ref{eq:lambchi2}) and require $|\lambda^\chi_2| < 5$.  Moreover, because $m_{B,N}\gg m_{H^\pm_{1,2}}$, the phenomenological results will not be sensitivity to the values of $m_{H^\pm_{1,2}}$.  In the numerical analysis, we fix $m_{H^\pm_1}=120$~GeV and $m_{H^\pm_2}=320$~GeV, which satisfy the constraints from the $T$ and $S$ parameters.  In addition,  the direct search bounds from colliders on $m_{H_I, A_I}$ are not very strict, and the bound on $m_{H^\pm_I}$, reinterpreted from the SUSY search at LEP, is $m_{H^\pm_I}> 70-90$~GeV ~\cite{Pierce:2007ut,Lundstrom:2008ai,Merchand:2019bod,Belanger:2021lwd}.  

  As alluded to before, the current upper bounds on the BRs of $t\to c(h, Z)$ and $h\to \mu \tau$ are less than $10^{-3}, 5\times 10^{-4}$, and $1.5\times 10^{-3}$, respectively.  In order to study the model predictions for these modes while avoiding the parameter space that is beyond the HL-LHC sensitivities, we further set low bounds on them when scanning the viable parameters.  The ranges  of the physical processes used to confine our parameter scan are summarized as follows:
 \begin{align}
 &  BR(t\to c h) \in (10^{-5},\, 10^{-3})\,, ~ BR(t\to c Z) \in (0.2,\, 5) \times 10^{-4}\,, ~ BR(h\to ee) < 3.6 \times 10^{-4}\,, \nonumber \\
 & BR(h\to e \mu) < 6.1\times 10^{-5}\,, ~BR(h\to \mu \tau) > 0.5\times  10^{-4}\,, ~ BR(\tau\to e\gamma) <  3.3 \times 10^{-8} \,, \nonumber\\
 & BR(\tau\to \mu\gamma) <  4.2 \times 10^{-8}\,, ~ a_\mu \in (1, \, 5) \times 10^{-9}\,, ~ a_e \in (-13,\,8)\times 10^{-13}\,. \label{eq:bounds}
 \end{align} 
Since the resulting $BR(h\to \mu \tau)$ is much smaller than the current upper limit, we only use the low bound to constrain the parameters. We note that the ranges of $a_\mu$ and $a_e$ are taken in such a way that $a_\mu$ can be positive and around $O(10^{-9})$, and the obtained values of $a_e$ are sufficiently wide so that the current experimental results shown in Eq.~(\ref{eq:Dae}) can be covered.

\subsection{ Numerical analysis and discussions}

In this subsection, we discuss the numerical results for the BRs of $t\to c (h, \gamma, Z)$, $\tau \to (e, \mu) \gamma$, $h\to \ell \ell'$, and lepton $g-2$ when the parameter ranges given in Eqs.~(\ref{eq:limits}) and (\ref{eq:bounds}) are used.  We first note that because of the constraint of $D$-meson mixing, FCNCs for $t\to c$ and $t\to u$ processes cannot be simultaneously enhanced up to the sensitivities of HL-LHC.  We focus on the $t\to c$ processes in the following analysis.  Although the BR of  $h\to \tau^- \tau^+$ in the SM is at the percent level, we here only exhibit the purely new physics effects in the numerical analysis.

\subsubsection{ Top-FCNCs}

\begin{figure}[phtb]
\begin{center}
\includegraphics[scale=0.5]{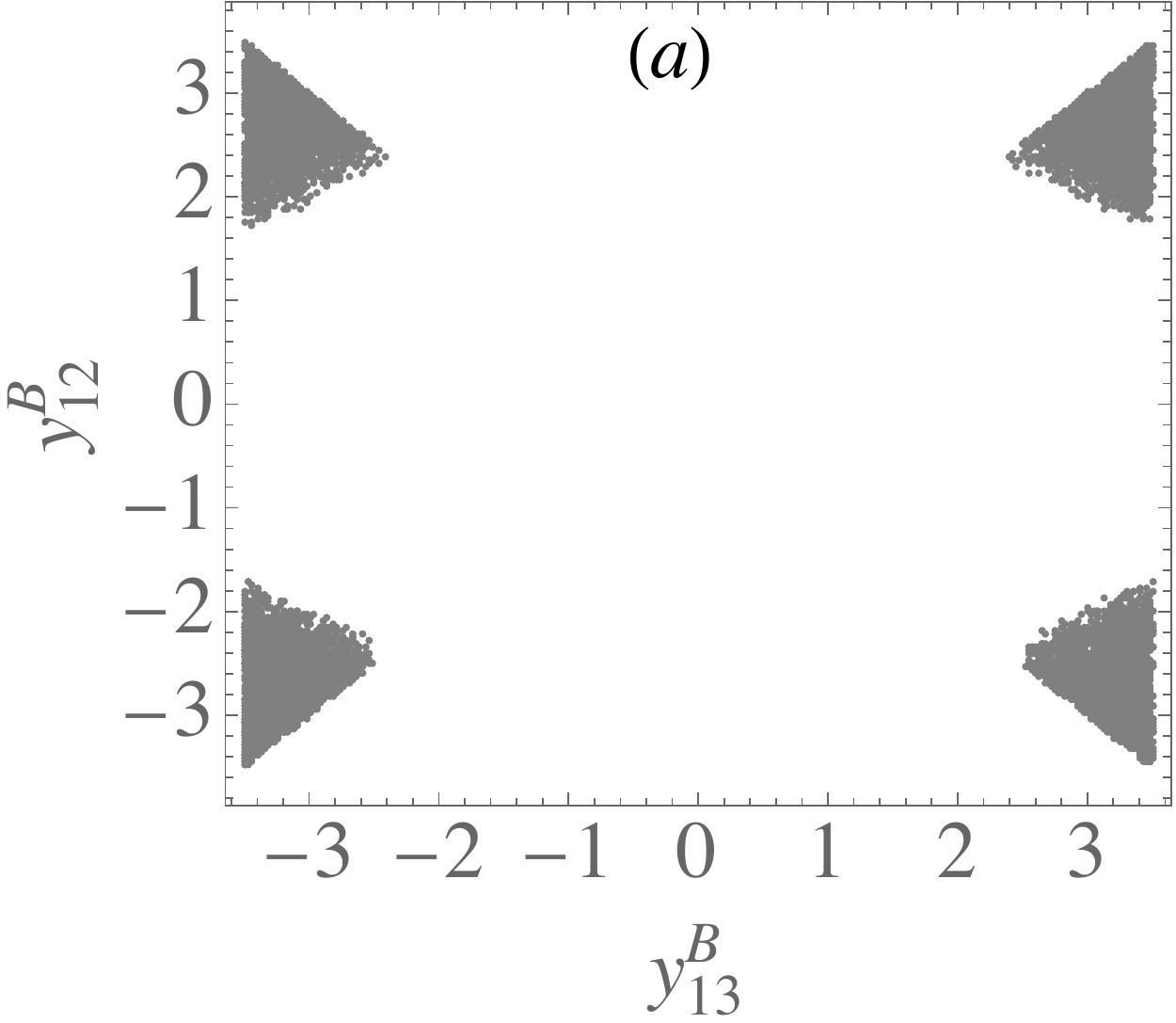}
\includegraphics[scale=0.5]{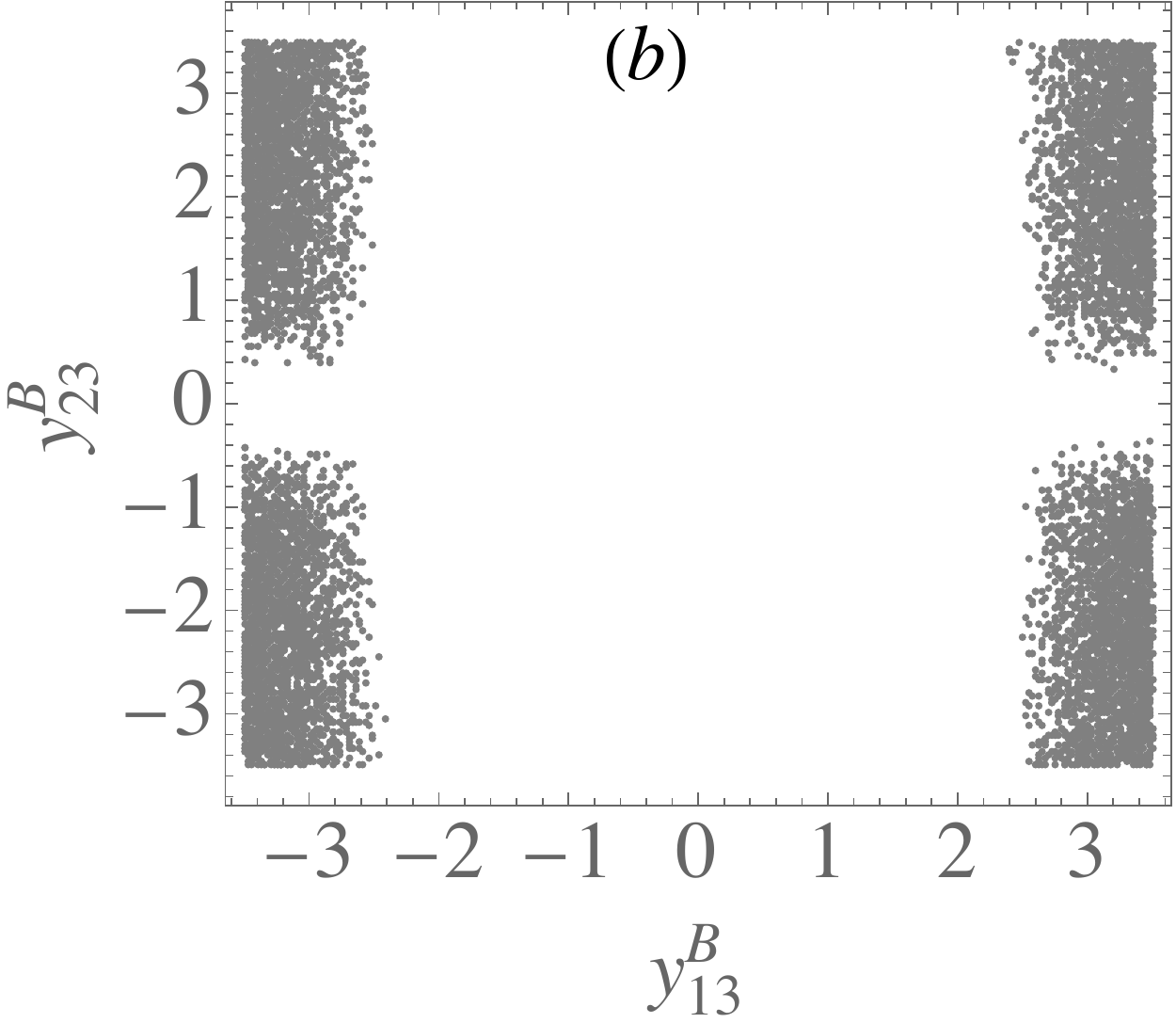}
\includegraphics[scale=0.5]{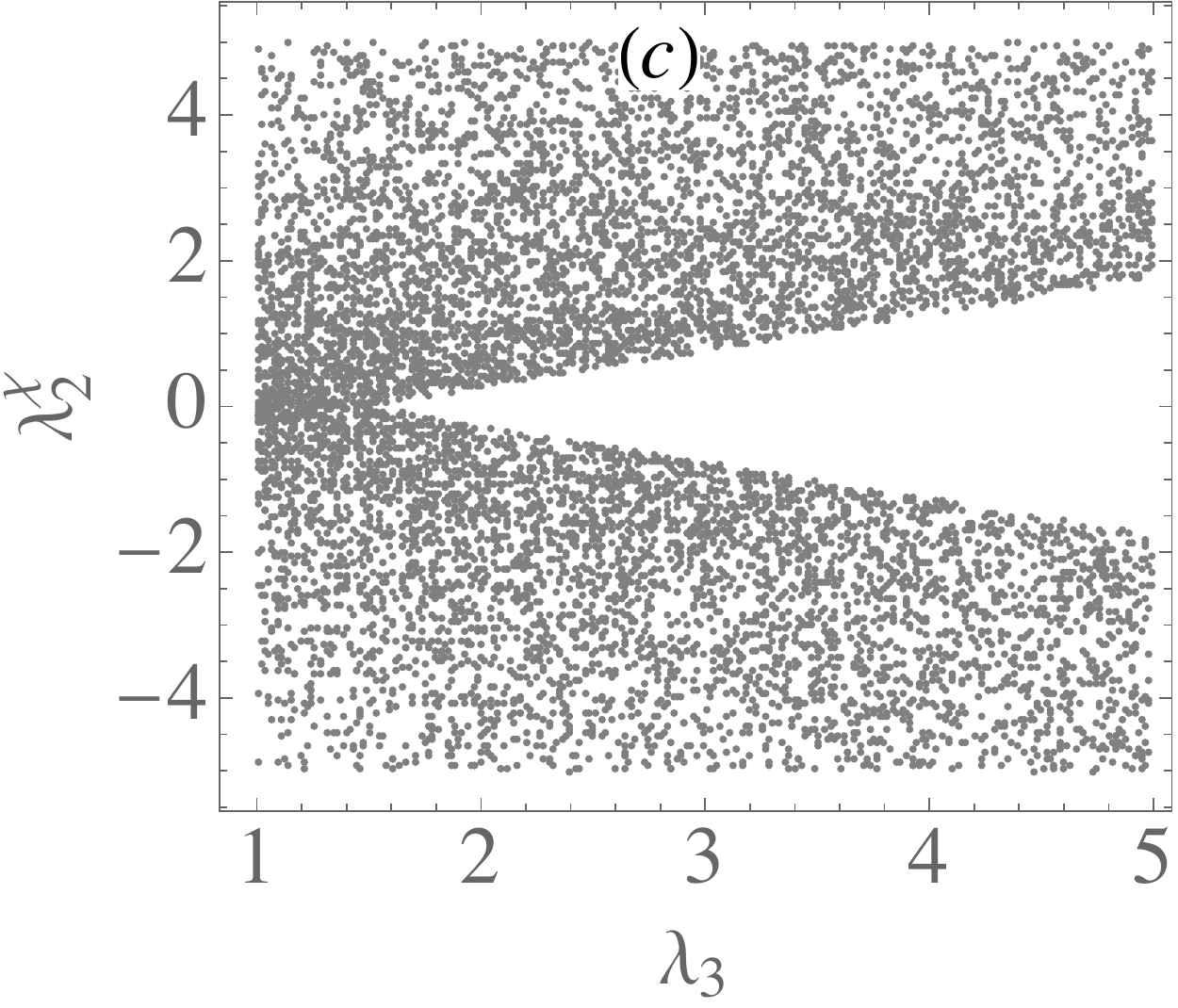}
\includegraphics[scale=0.5]{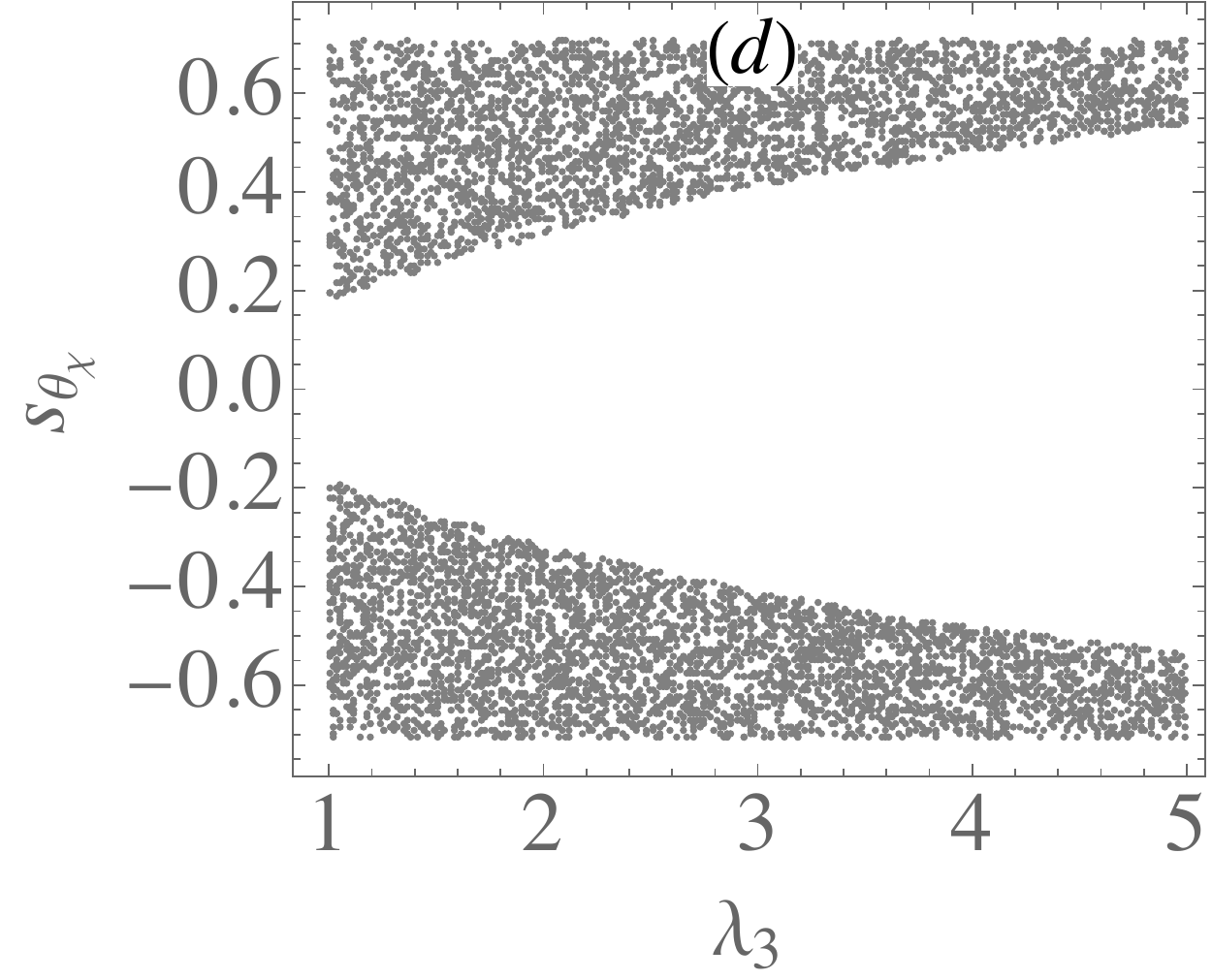}
 \caption{ Scatter plots for the allowed parameter space projected in the planes of:
 (a) $y^B_{13}-y^{B}_{12}$,  (b) $y^B_{13}-y^{B}_{23}$, (c) $\lambda_3-\lambda^\chi_2$, and (d) $\lambda_3-s_{\theta_\chi}$.   }
\label{fig:top_para}
\end{center}
\end{figure}

 To see the distributions of parameters that fit the ranges of $t\to c (h, Z)$ given in Eq.~(\ref{eq:bounds}), we show the scatter plots for the correlations of parameters in Fig.~\ref{fig:top_para}, where plots (a), (b), and (c) denote are for the $y^B_{13}-y^{B}_{12}$, $y^B_{13}-y^{B}_{23}$, and $\lambda_3-\lambda^\chi_2$ planes, respectively.  It can be found that to reach the BRs of top-FCNCs at the level of  ${\cal O}(10^{-5})$, the Yukawa couplings $y^{B}_{12,13}$ have to be large and  are restricted in a narrow range of $|y^{B}_{12,13}| \lesssim  (2.6, 3.5)$, whereas $y^{B}_{23}$ is wider and $|y^{B}_{23}| \lesssim  (0.3, 3.5)$.  From Eqs.~(\ref{eq:tqh}) and (\ref{eq:tqZ}), it is known that  $BR(t\to ch/cZ)$ do not vanish when $s_{\theta_\chi}=0$. However,  because we require that all parameters have to fit the chosen ranges in Eq.~(\ref{eq:bounds}) for the top decays, the values of $s_{\theta_\chi}$ may influence their BRs.  Thus, we show the correlation between $s_{\theta_\chi}$ and $\lambda_3$ in Fig.~\ref{fig:top_para}(d). 

\begin{figure}[tbph]
\begin{center}
\includegraphics[scale=0.5]{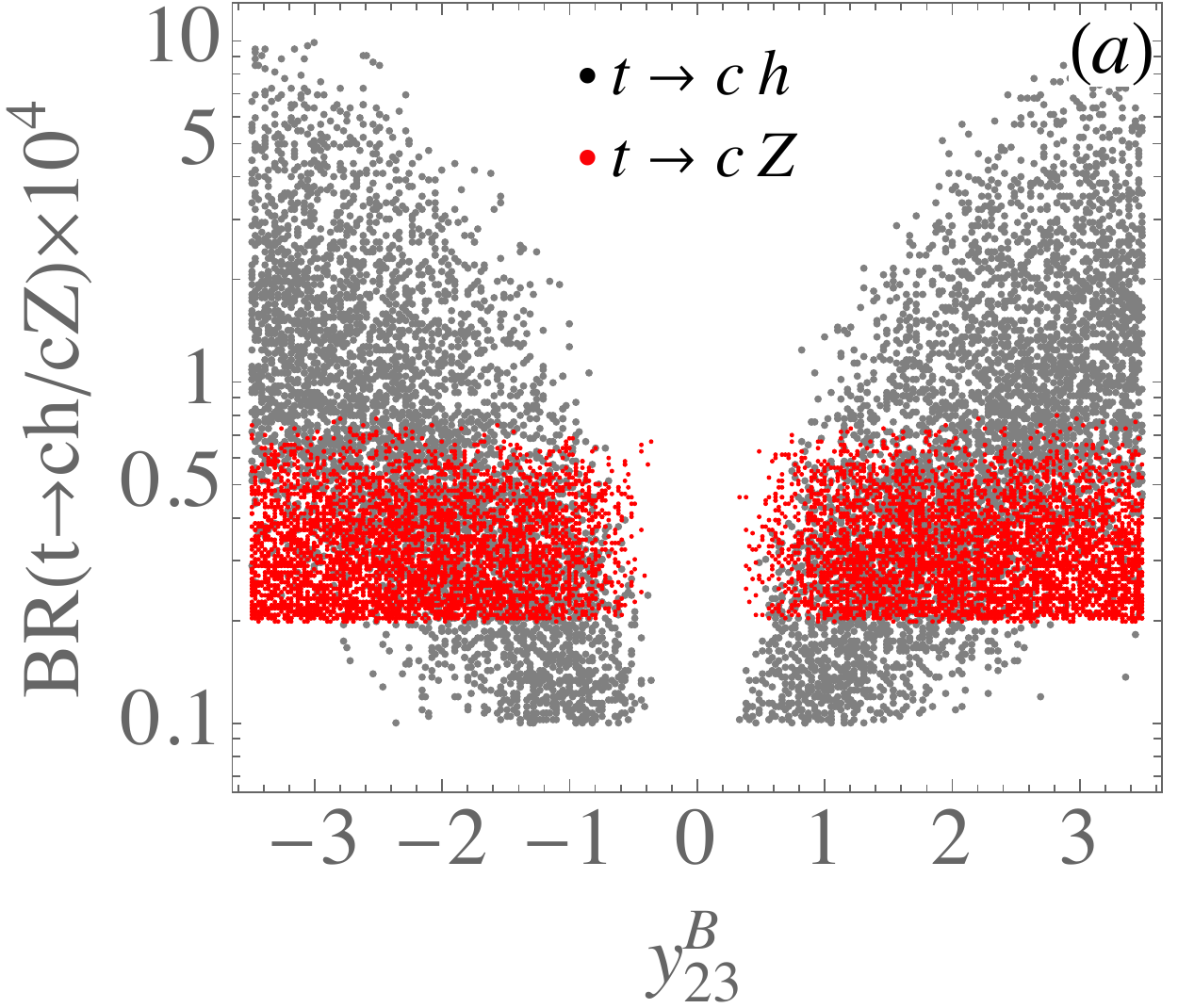}
\includegraphics[scale=0.5]{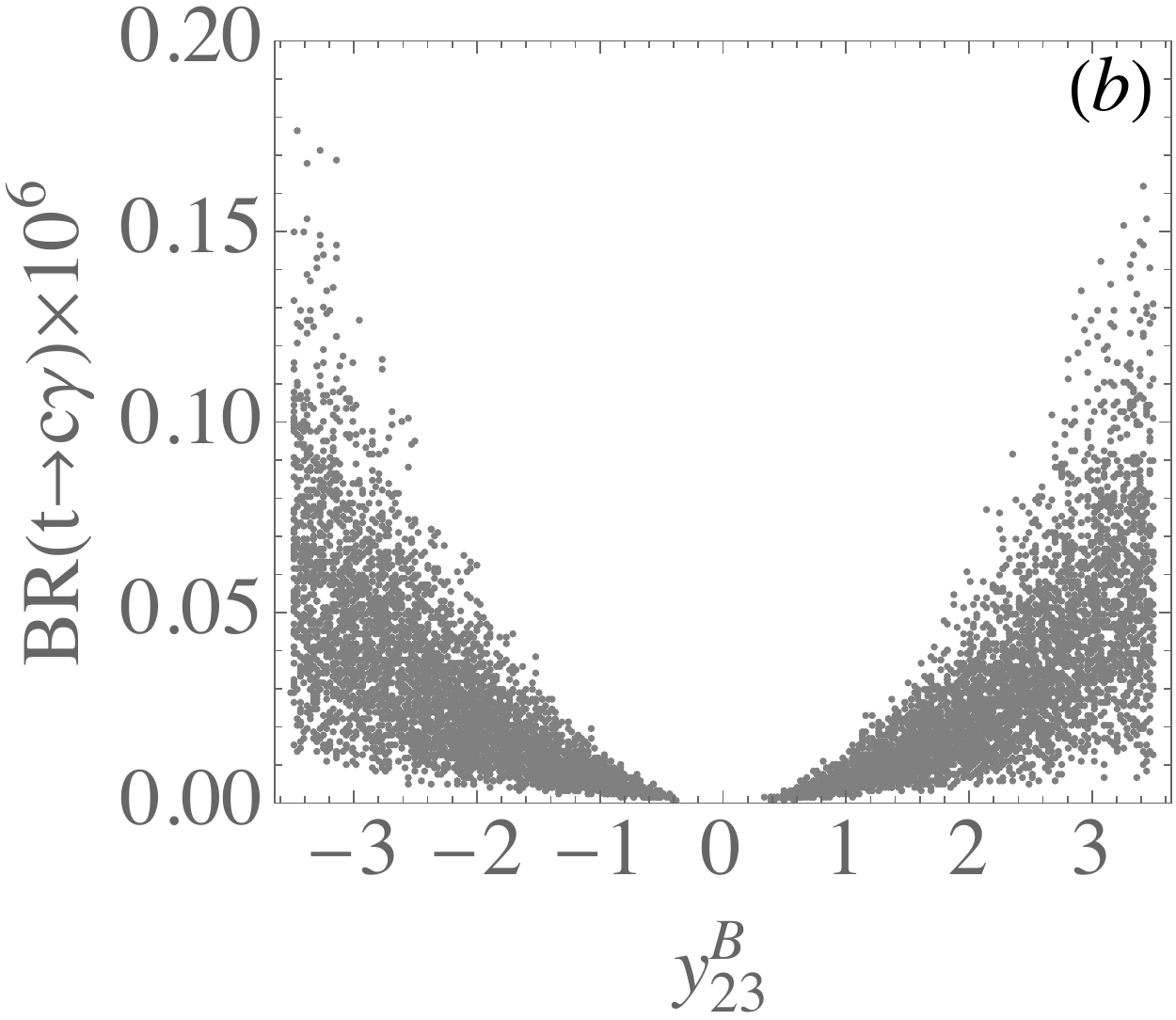} 
\includegraphics[scale=0.5]{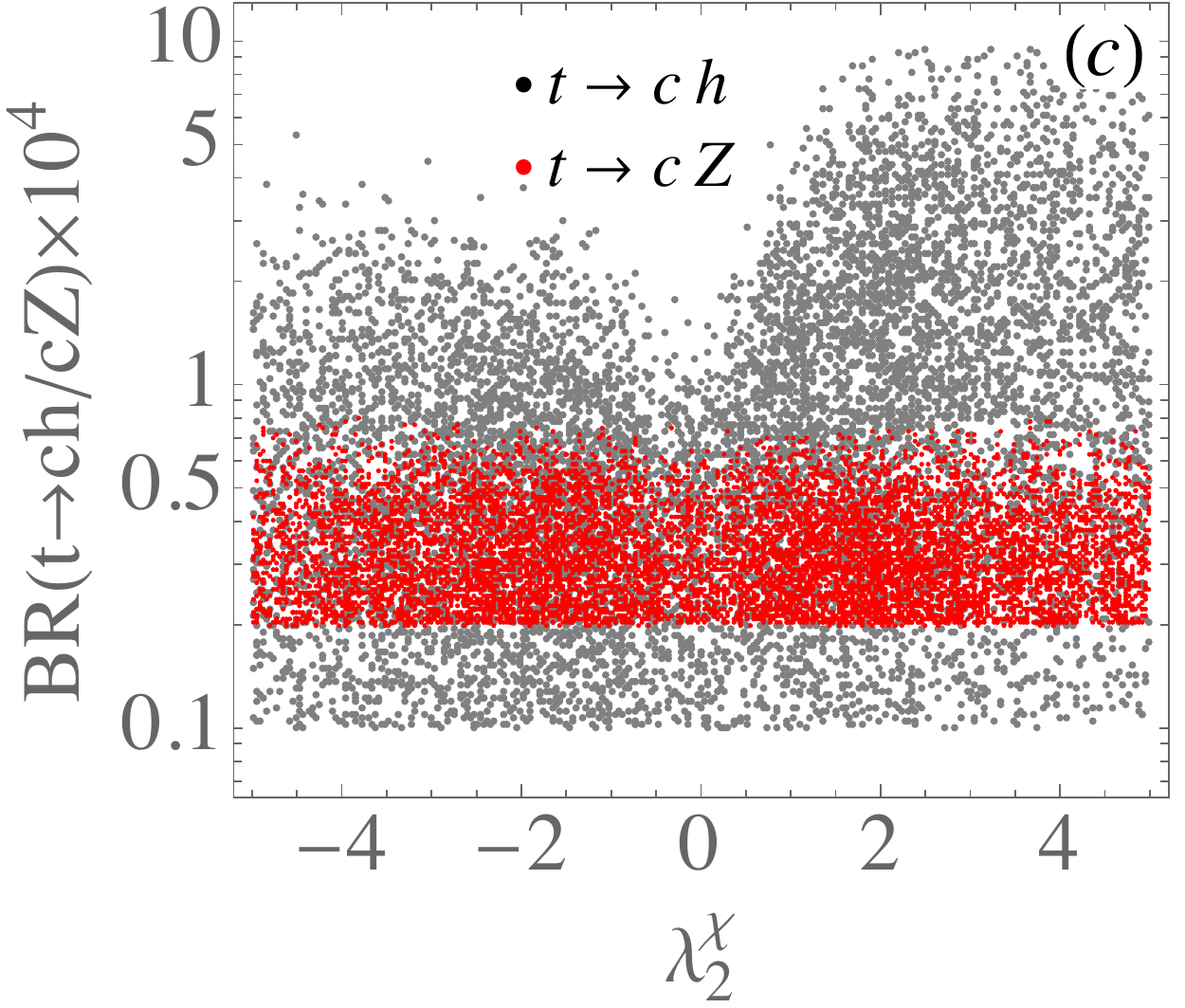}
\includegraphics[scale=0.5]{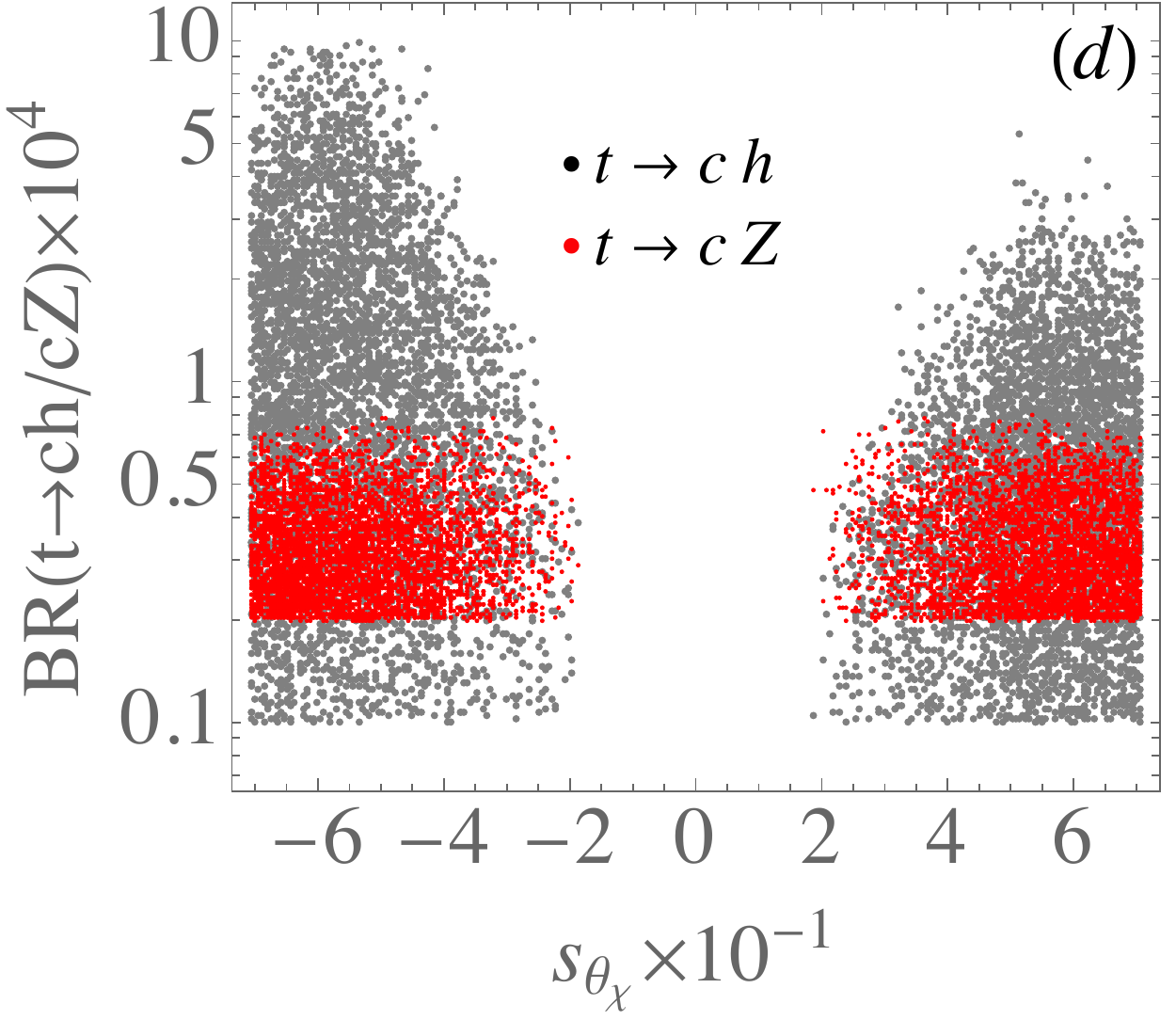}
 \caption{ Scatter plots for $BR(t\to ch/cZ)$ as a function of 
 (a) $y^B_{23}$,  (c) $\lambda^\chi_2$, and (d) $s_{\theta_\chi}$.  Plot (b) is a scatter plot for the correlation between $BR(t\to c\gamma)$ and $y^{B}_{23}$.   }
\label{fig:top_FCNC}
\end{center}
\end{figure}

After obtaining the available  parameter space, we show the BRs of the top-FCNC decays as a function of different parameters in Fig.~\ref{fig:top_FCNC}. Since $y^B_{12,13}$ are constrained within a narrow range, we only exhibit the $y^B_{23}$ dependence for the Yukawa couplings. From Fig.~\ref{fig:top_FCNC}(a), it is seen that $BR(t\to c h)\sim {\cal O}(10^{-4})$ and $BR(t\to c Z)\sim {\cal O}(10^{-5})$ can be achieved, with the former be possibly more than one order of magnitude larger than the latter. The results can be understood as follows.  From Eqs.~(\ref{eq:tqh}) and (\ref{eq:tqZ}), the main different factors in $t\to ch$ and $t\to cZ$ can be expressed  as $[ \lambda^h_{12}(c^2_{\theta_\chi}-s^2_{\theta_\chi} )-\lambda^h_{22}s_{\theta_\chi} c_{\theta_\chi} ]  v/m_B$ and $g s^2_W Q_B/c_W\sim 0.01$, respectively. Thus, even with $c_{\theta_\chi}\sim s_{\theta_\chi}$ or  $s_{\theta_\chi}\sim 0$, the result  $BR(t\to ch) \gg BR(t\to cZ)$ is expected  when $\lambda^h_{22} \sim {\cal O}(1)$ or  $\lambda^h_{12}\sim {\cal O}(1)$. In addition, since we have taken $f^B_{21,22}\approx 0$ due to the constraints of $\Delta B=2$ and $\Delta D=2$ processes, it is seen that $t\to cZ$ becomes independent of $y^B_{23}$. Although $t\to cZ$ is not sensitive to $y^{B}_{23}$, the excluded region for $|y^{B}_{23}|\lesssim 0.4$ shown in Fig.~\ref{fig:top_FCNC}(a) is from the requirement that the relevant parameters have to satisfy the taken ranges of the top decays shown in Eq.~(\ref{eq:bounds}).  

 From Fig.~\ref{fig:top_FCNC}(b), we see that the BR for the $t\to c \gamma$ decay in the model is far below $10^{-6}$. According to Eq.~(\ref{eq:tqgamma}), as alluded to earlier, a suppression effect arises from the cancellation between $H^\pm_1$ and $H^\pm_{2}$.  If we take $m^2_{H^\pm_i}/m^2_B \approx 0$ in the loop integrals $ J^{1,2}_\gamma$, the dominant $B^\gamma_L$ vanishes.  We note that $B^\gamma_R\approx 0$ because $y^B_{21,22}\approx 0$ is applied. Therefore, using the assumed values of $m_{H^\pm_i}$, $B^\gamma_L$ is suppressed even though it does not completely vanish.  Hence, $BR(t\to c \gamma)$ cannot be possibly enhanced up to ${\cal O}(10^{-6})$.  In addition, according to Eq.~(\ref{eq:tqgluon}), we see that the BR for $t\to c g$ can reach ${\cal O}(10^{-6})$, which is still much smaller than the upper limit measured by ATLAS with $BR^{\rm exp}(t\to c g) < 3.7 \times 10^{-4}$~\cite{ATLAS:2021amo}.
 
 It is known that in addition to the Yukawa couplings, the Higgs trilinear couplings to the charged Higgses $\lambda^h_{ij}$ defined in Eq.~(\ref{eq:Higgs_triple}) play an important role in the $t\to c h$ decay.  Since $\lambda^h_{ij}$ consist of $\lambda_3$, $\lambda^\chi_2$, and $s_{\theta_\chi}$, to see the dependence of quartic scalar couplings in the decay, the correlation between $BR(t\to ch/cZ)$ and  $\lambda^\chi_2$, defined in Eq.~(\ref{eq:lambchi2}), is shown in Fig.~\ref{fig:top_FCNC}(c). The resulting dependence pattern for $\lambda_3$ is similar.  Since $t\to cZ$ is not related to $\lambda_3$ and $\lambda^\chi_2$, it is seen that $BR(t\to cZ)$ is insensitive to the quartic scalar couplings.  The dependence of $BR(t\to ch/cZ)$ on $s_{\theta_\chi}$ is shown in Fig.~\ref{fig:top_FCNC}(d).  As stated earlier, the dominant effects in $t\to ch$ are associated with $c^2_{\theta_\chi}-s^2_{\theta_\chi}$ and $s_{\theta_\chi} c_{\theta_\chi}$, whereas $t\to cZ$ is not sensitive to $s_{\theta_\chi}$ because the decay amplitude results in  $c^2_{\theta_\chi} + s^2_{\theta_\chi}=1$.   Since  the parameters need to fit the  values given in Eq.~(\ref{eq:bounds}),  we thus obtain a null result of $t\to cZ$ in the excluded region of $|s_{\theta_\chi}| \lesssim 0.2$.

\subsubsection{ Leptonic Higgs decays, radiative LFVs, and lepton $g-2$}

\begin{figure}[phtb]
\begin{center}
\includegraphics[scale=0.5]{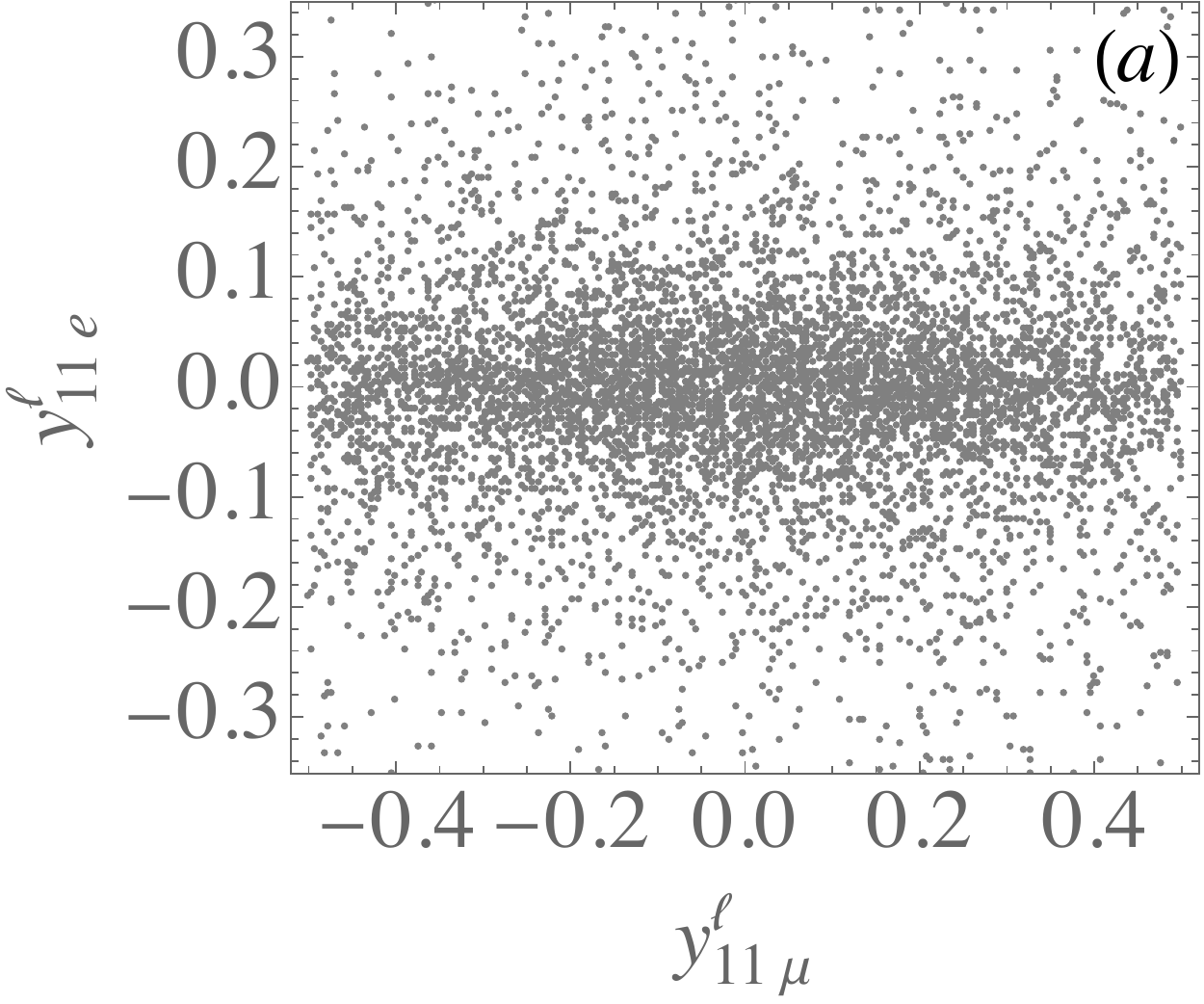}
\includegraphics[scale=0.5]{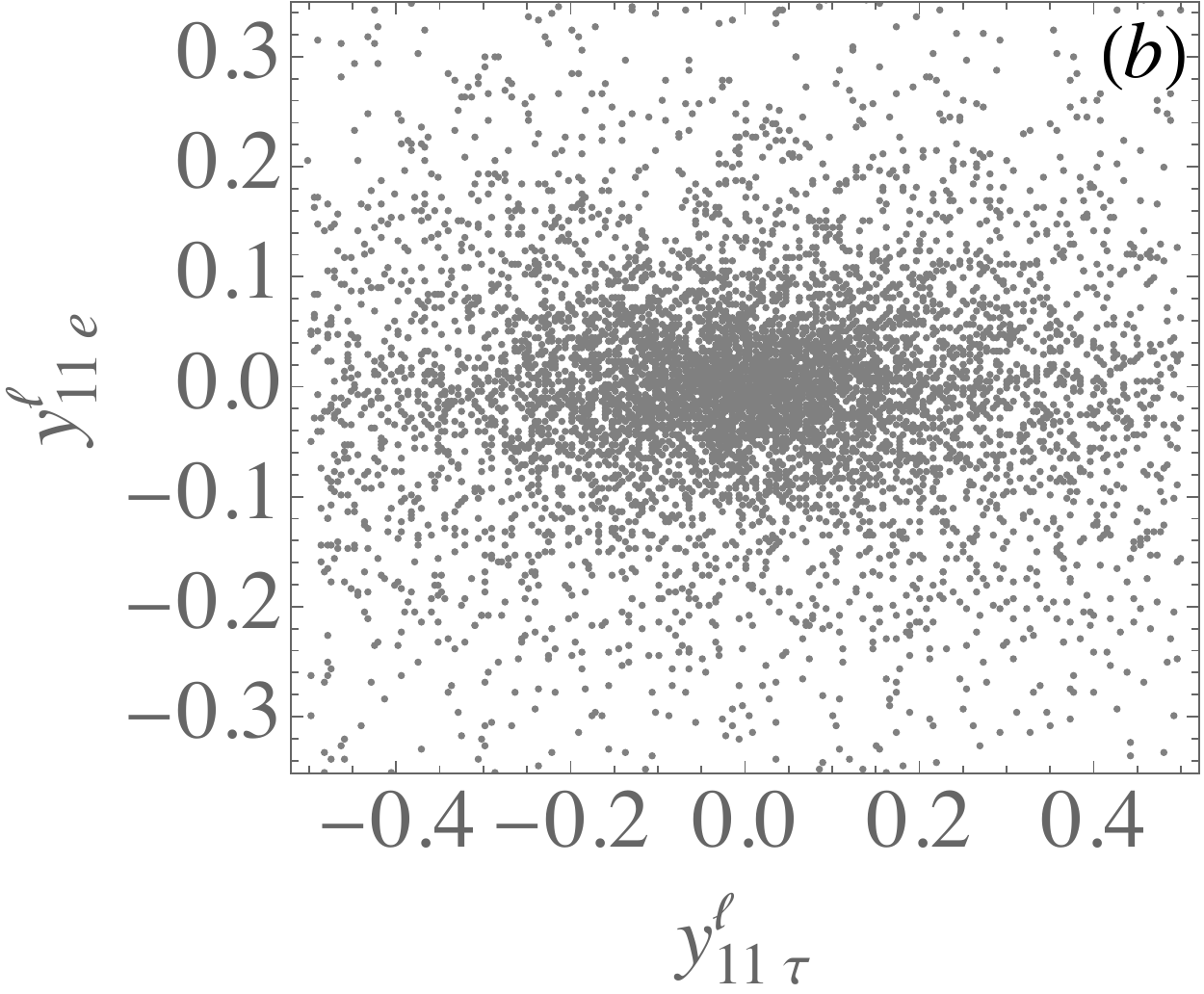}
\includegraphics[scale=0.5]{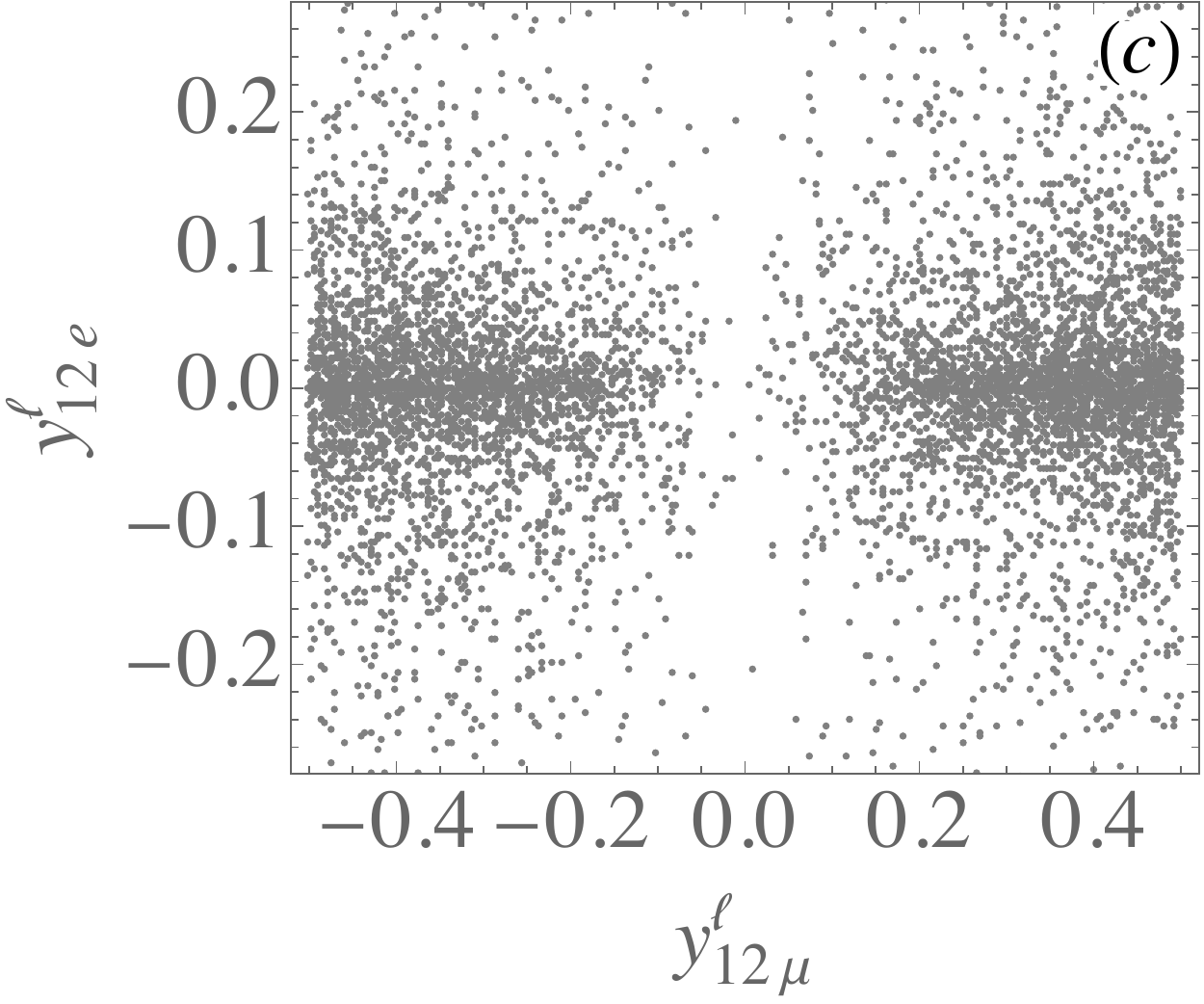}
\includegraphics[scale=0.5]{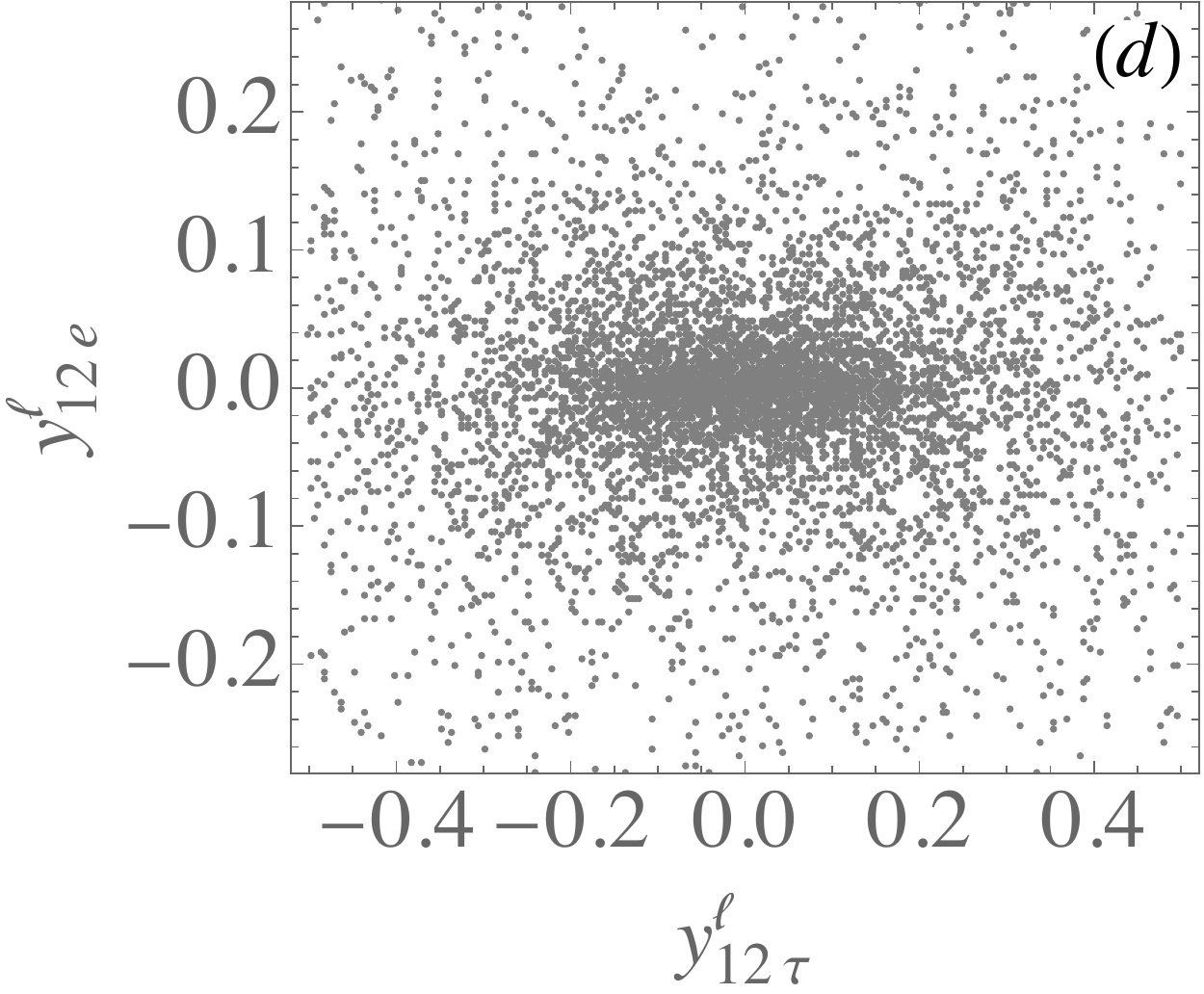}
 \caption{ Scatter plots for the correlations among ${\bf y}^\ell_{11}$ and ${\bf y}^\ell_{12}$, where the values of  parameters are selected to fit the regions of physical observables in Eq.~(\ref{eq:bounds}).   }
\label{fig:LFV_para}
\end{center}
\end{figure}

In addition to the Higgs trilinear couplings $\lambda^h_{ij}$, the processes $h\to \ell_i \ell_j$ as well as  $\ell \to \ell' \gamma$ and lepton $g-2$ are sensitive to the Yukawa couplings ${\bf y}^\ell_{1k, 2k}$ and $m_{N_k}$.  To see the distribution patterns of  ${\bf y}^\ell_{1k,2k}$ when the constraints in Eq.~(\ref{eq:bounds}) are satisfied, we show various correlations of the Yukawa couplings in Fig.~\ref{fig:LFV_para}. Since  the correlations among ${\bf y}^{\ell}_{2k}$ are similar to those among ${\bf y}^\ell_{1k}$, we only exhibit the scatter plots for the correlations of Yukawa couplings, $y^{\ell}_{11e}-y^{\ell}_{11\mu}$, $y^{\ell}_{11e}-y^{\ell}_{11\tau}$, $y^{\ell}_{12e}-y^{\ell}_{12\mu}$, and $y^{\ell}_{12e}-y^{\ell}_{12\tau}$, in Figs.~\ref{fig:LFV_para}(a)-(d), respectively.  From the plots, it can be seen that  $|y^\ell_{11e,12e}|$  have denser sample points below 0.1, whereas the allowed values of $y^\ell_{11\mu,12\mu}$ and $y^\ell_{11\mu, 12\tau}$ are wider.  It is found that the correlations of $\lambda_3-\lambda^\chi_2$ and $\lambda_3-s_{\theta_\chi}$ in leptonic decays are close to the results shown in Figs.~\ref{fig:top_para}(c) and (d); therefore, we skip the related plots. 

\begin{figure}[phtb]
\begin{center}
\includegraphics[scale=0.5]{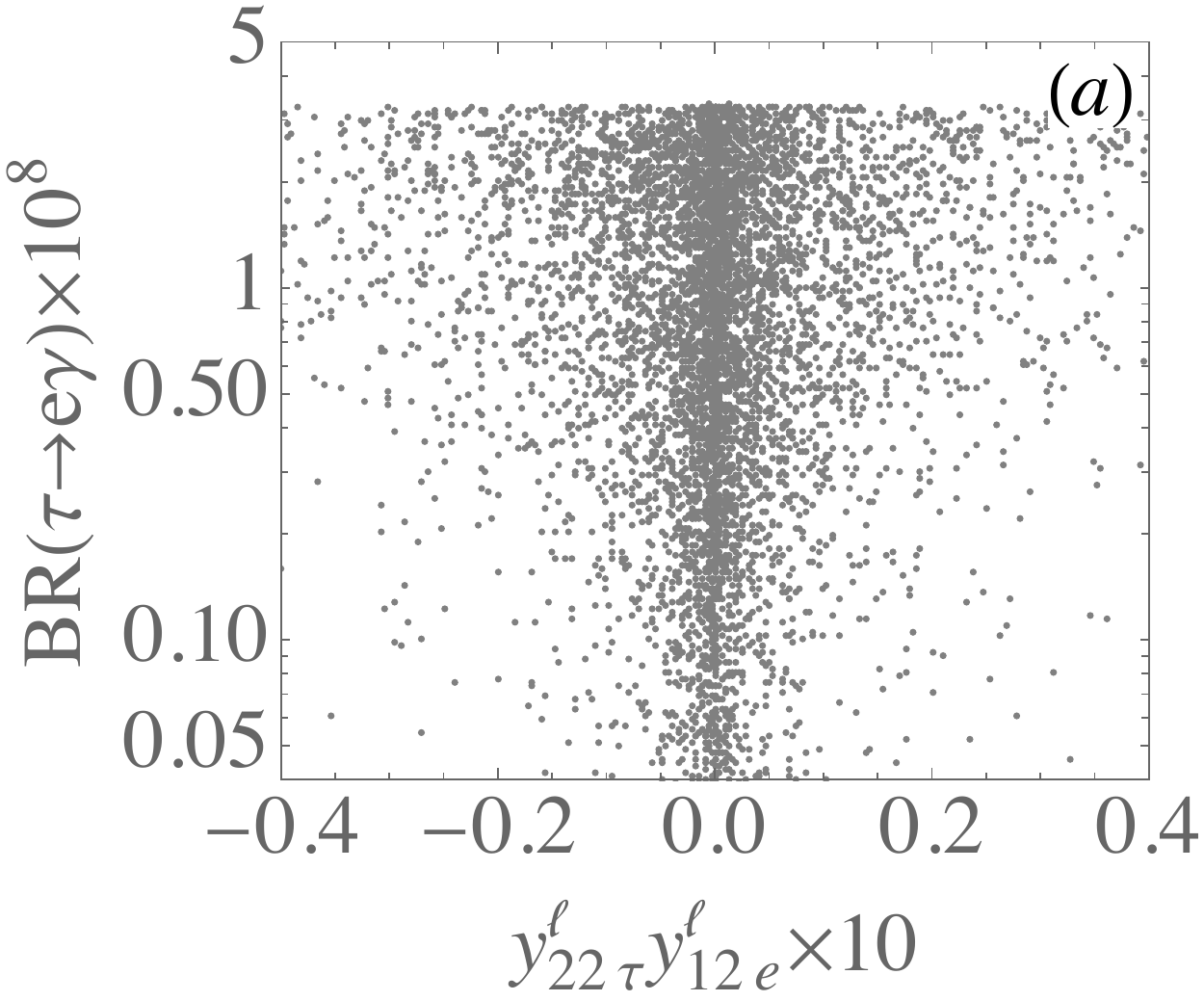}
\includegraphics[scale=0.5]{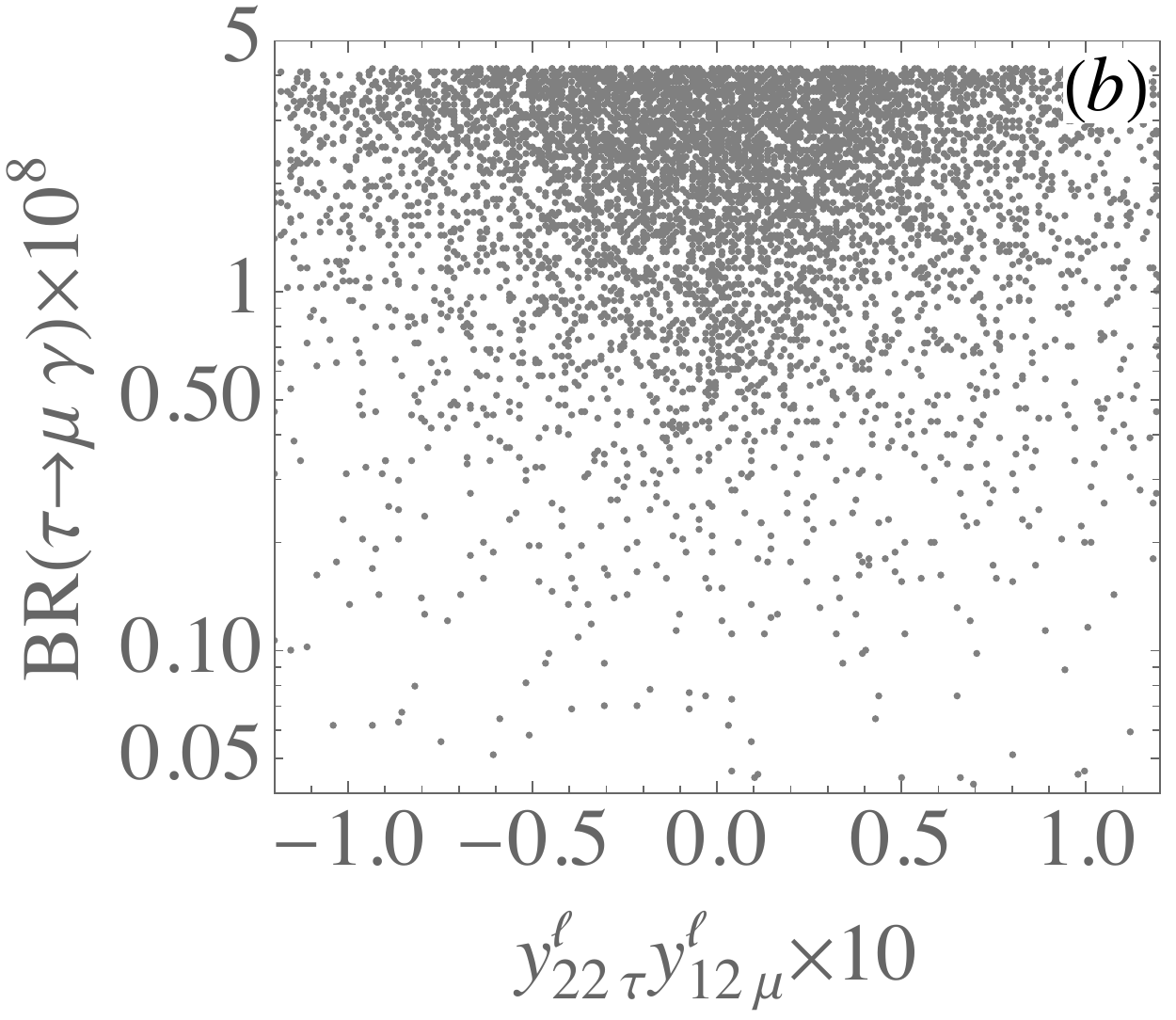}
\includegraphics[scale=0.5]{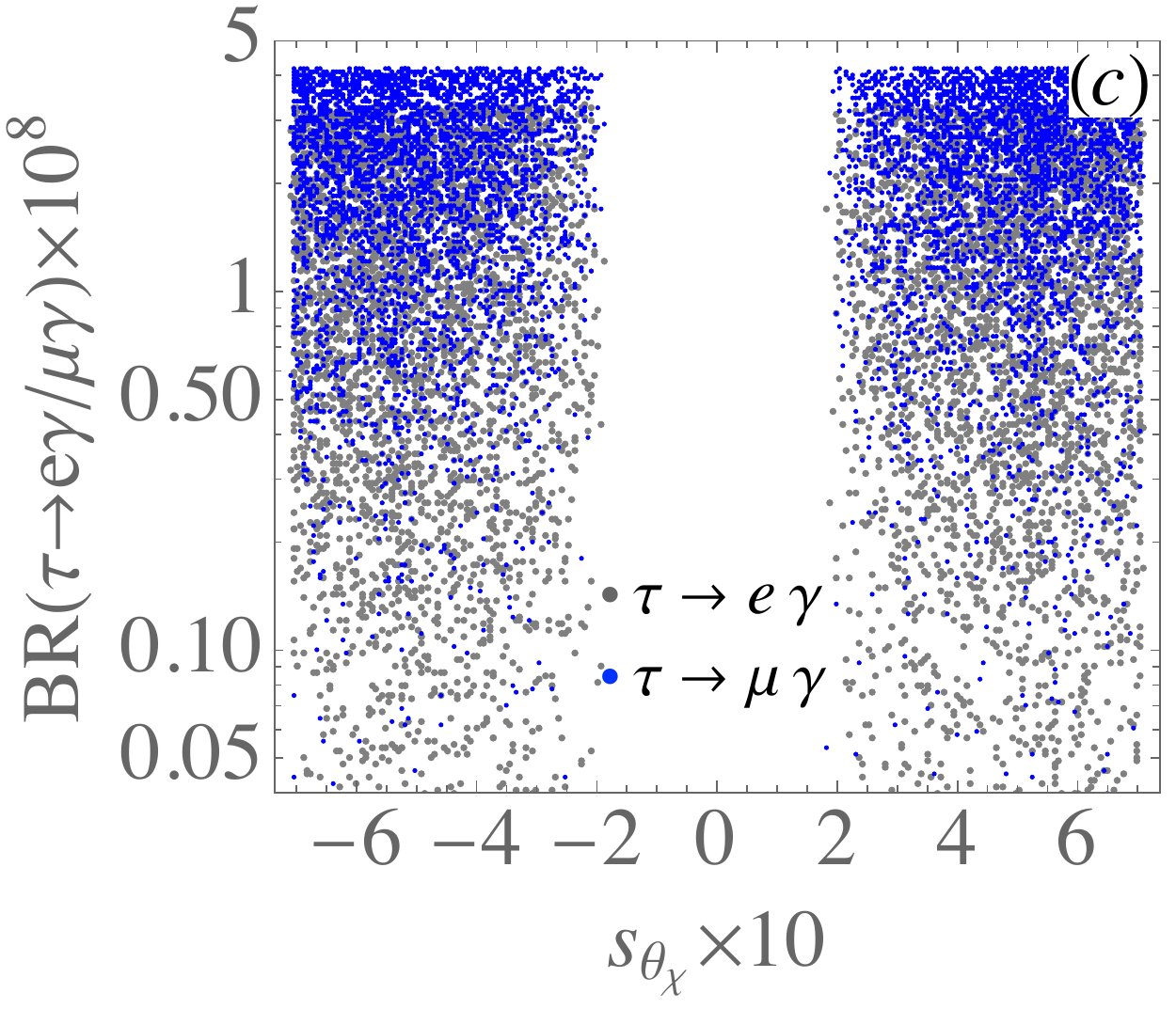}
\includegraphics[scale=0.5]{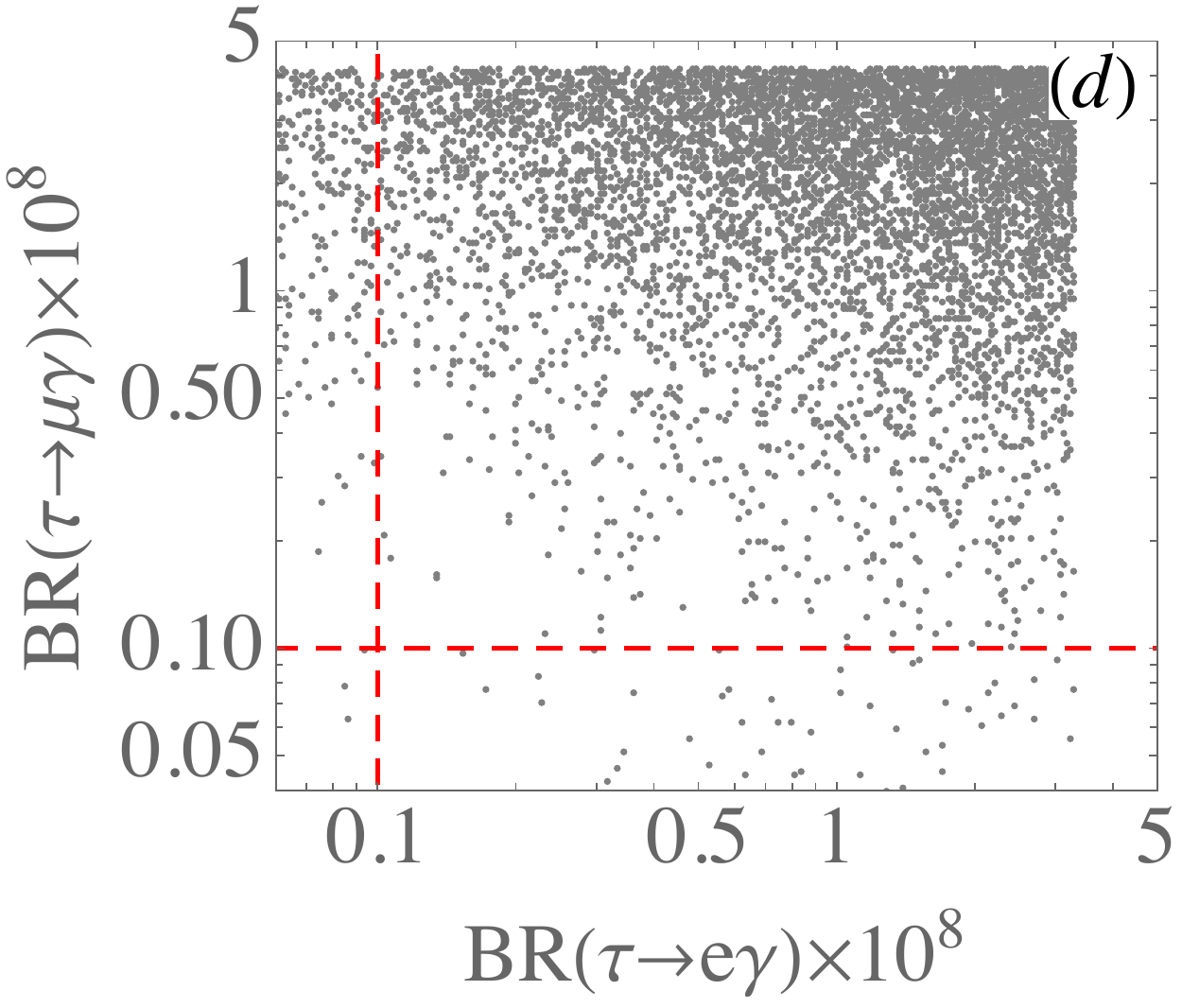}
 \caption{ Scatter plots for $BR(\tau\to \ell' \gamma)$ as a function of (a) $y^\ell_{22\tau} y^\ell_{12e}$ ($\ell'=e)$,  (b) $y^\ell_{22\tau} y^\ell_{12\mu}$ ($\ell=\mu$), and  (c) $s_{\theta_\chi}$.  Plot (d) shows the correlation between $BR(\tau\to e\gamma)$ and $BR(\tau\to \mu \gamma)$, where the dashed lines are the sensitivity levels of Belle II.    }
\label{fig:LFV}
\end{center}
\end{figure}

To see how $BR(\tau \to \ell' \gamma)$ are sensitive to the Yukawa couplings, we show the representative  dependence of $y^\ell_{22\tau} y^\ell_{12e}$ and $y^\ell_{22\tau} y^\ell_{12\mu}$  for $\tau \to e \gamma$ and $\tau \to \mu \gamma$ in Figs.~\ref{fig:LFV}(a) and (b), respectively, where the product of Yukawa couplings is taken from $C^\gamma_{L \ell' \ell}$ in Eq.~(\ref{eq:C_litoljga}) and other possible products of Yukawa couplings, such as $y^\ell_{21\tau} y^\ell_{11e}$ and $y^\ell_{21\tau} y^\ell_{11\mu}$, give a similar pattern for each radiative $\tau$ decay.  According to the results in Fig.~\ref{fig:LFV_para}(c), most sampling points for $|y^\ell_{12e}|$ are located at the values less than 0.1, whereas the distribution of $|y^\ell_{12\mu}|$ is wider and the sampling points at around $0.1$ are suppressed. Hence, the differences of patterns in Figs.~\ref{fig:LFV}(a) and (b) can be understood using the results shown in Fig.~\ref{fig:LFV_para}.  The correlations with $s_{\theta_\chi}$ are shown in~Fig.~\ref{fig:LFV}(c), and the resulting pattern is similar to the $t\to ch/cZ$ decays in Fig.~\ref{fig:top_FCNC}, where $|s_{\theta_\chi}| \lesssim 0.2$ cannot fit the assumed ranges in Eq.~(\ref{eq:bounds}).  For the purpose of clarity, we show the correlation plot of $BR(\tau \to e \gamma)$ and $BR(\tau\to \mu\gamma)$ in Fig.~\ref{fig:LFV}(d), where the dashed lines denote the expected sensitivities of Belle II.  Since we use the current experimental upper limits to bound the free parameters, most sampling points are located around the regions close to the current upper limits.  The sampling points will predict lower $BR(\tau \to \ell'\gamma)$ when stricter bounds from experiments are obtained. 

\begin{figure}[tbph]
\begin{center}
\includegraphics[scale=0.5]{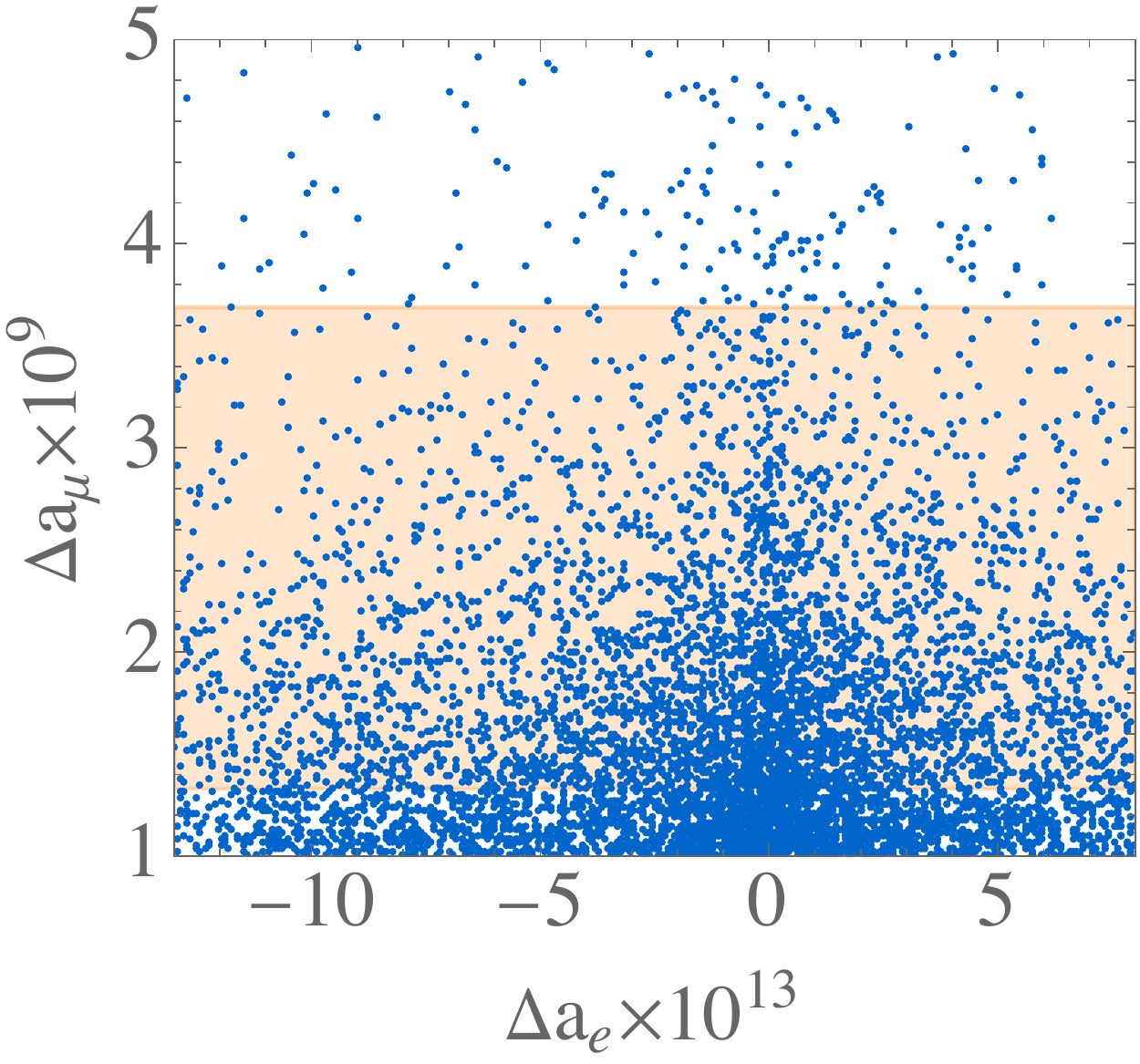}
 \caption{Correlation between electron and muon $(g-2)$'s in the model, where the light orange band denotes the data within $2\sigma$ errors taken from Eq.~(\ref{eq:Damu}). }
\label{fig:lepton_gm2}
\end{center}
\end{figure}

Using Eq.~(\ref{eq:lepton_gm2}) and the bounded parameter values obtained in radiative $\tau$-lepton decays, we can estimate the contributions to the lepton $(g-2)$'s. The correlation between $\Delta a_e$ and $\Delta a_\mu$ is shown in Fig.~\ref{fig:lepton_gm2}, where the band denotes $\Delta a^{\rm exp}_\mu$ in its $2\sigma$ range.  It is clearly seen that $\Delta a_\mu$ in the model can explain the muon $g-2$ anomaly, and the electron $g-2$ from the inert charged Higgses has the freedome to be either positive or negative~\cite{Dcruz:2022dao}. It is expected that with more precise measurement on the fine structure constant, e.g., from $\Delta a_e$, the relevant parameters can be further limited.  Since the Yukawa couplings involved in $\Delta a_e$ and $\Delta a_\mu$ are different, even the future data exhibit that $\Delta a_e$ is consistent with the SM prediction, $\Delta a_\mu \sim {\cal O}(10^{-9})$ can still be accommodated by the model. 

\begin{figure}[phtb]
\begin{center}
\includegraphics[scale=0.5]{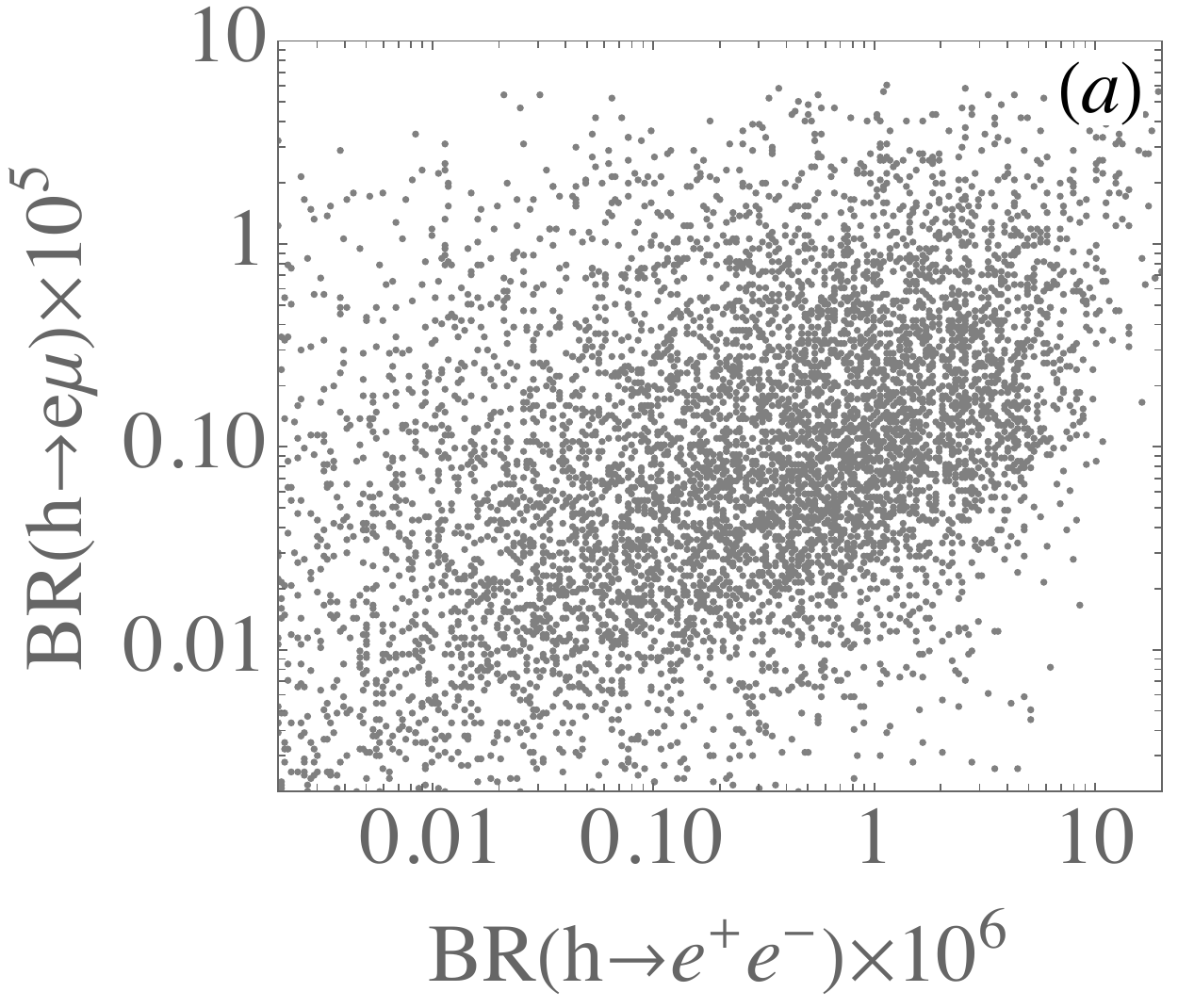}
\includegraphics[scale=0.5]{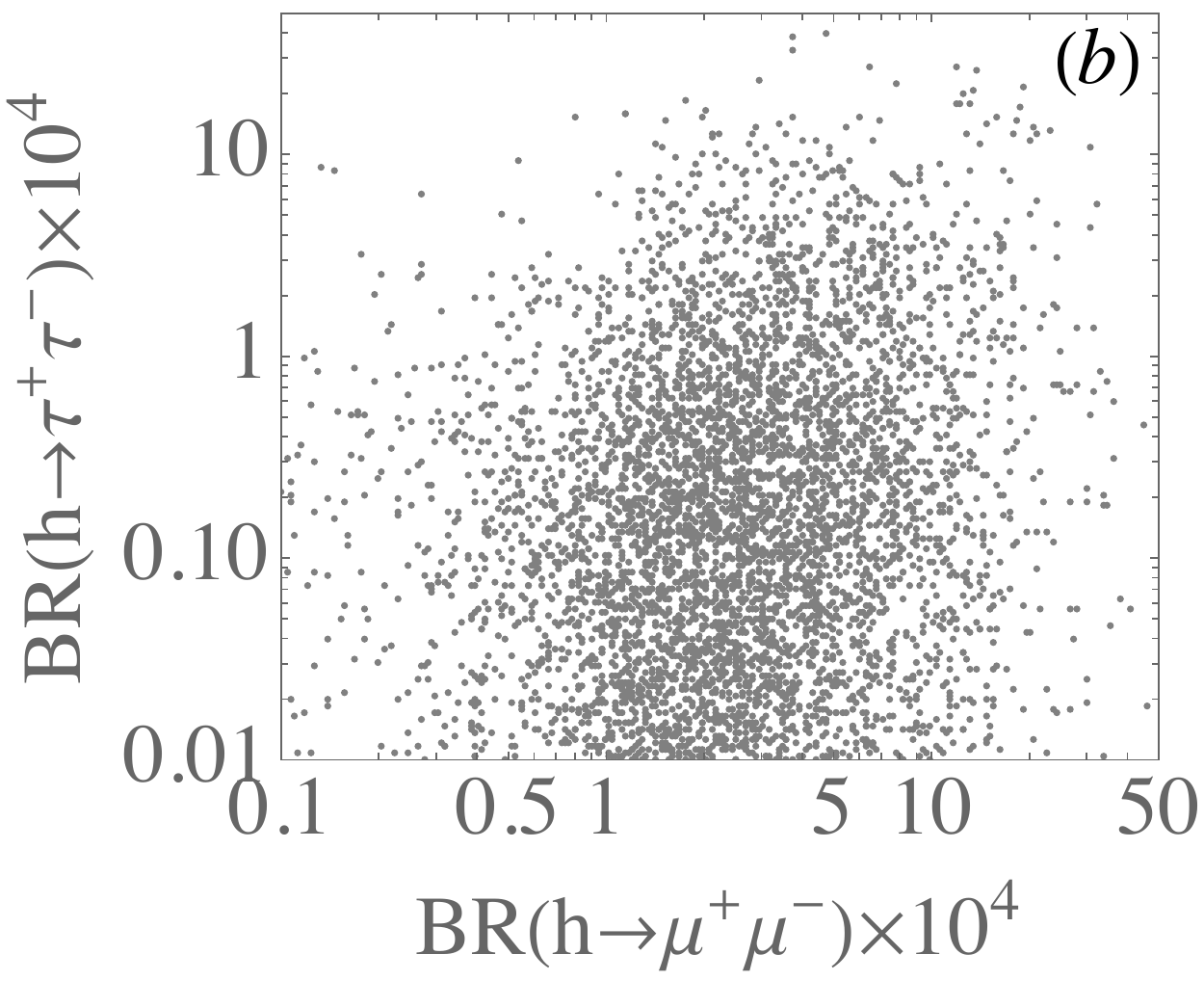}
\includegraphics[scale=0.5]{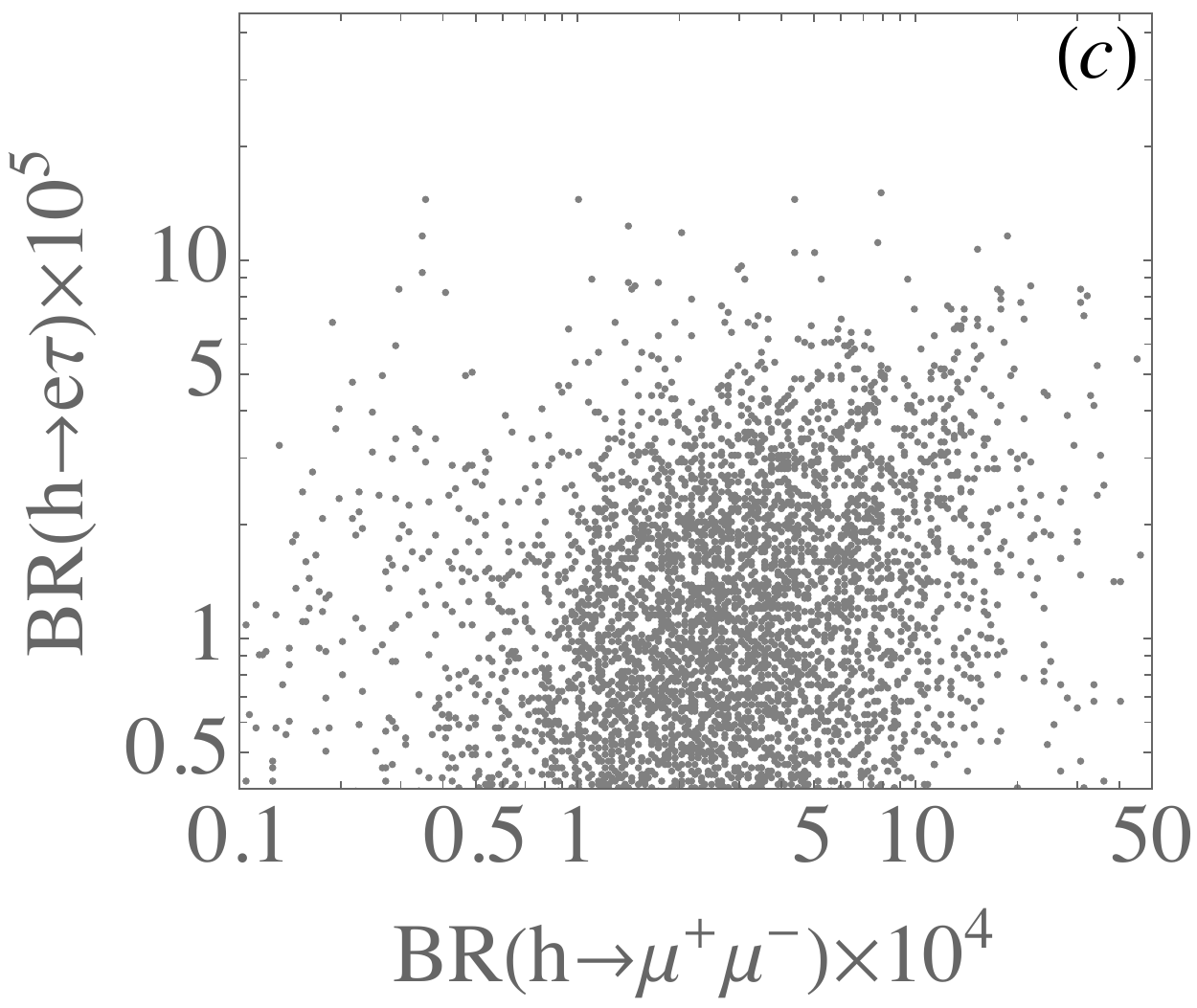}
\includegraphics[scale=0.5]{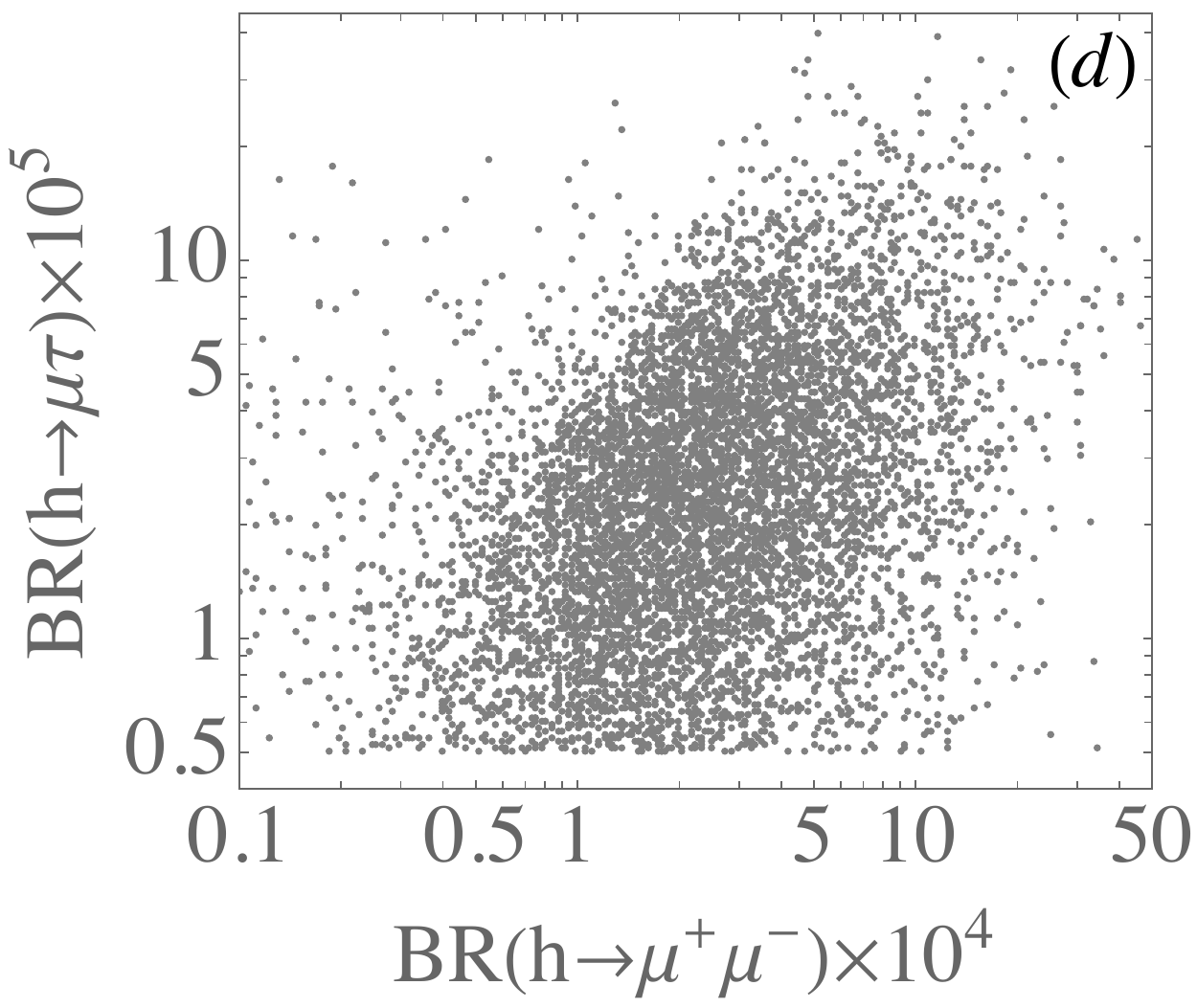}
 \caption{  Scatter plots for the correlations among  $BR(h\to \ell \ell')$. }
\label{fig:Lepton_Higgs}
\end{center}
\end{figure}

Next, we discuss the leptonic Higgs decays.  From Table~\ref{tab:hellellp}, it is known that the currently  strictest upper bounds on the $h\to \ell \ell'$ decays are  the $h\to e e/e\mu$  decays.  Combing with the constraints from the charged-lepton decays, the resulting $BR(h\to ee)$ and $BR(h\to e\mu)$ are shown in Fig.~\ref{fig:Lepton_Higgs}(a). It is seen that the BRs of both decays fall preferably at around ${\cal O}(10^{-6})$ though the values of ${\cal O}(10^{-5})$ can be achieved as well.  From Fig.~\ref{fig:Lepton_Higgs}(b), it can be found that the allowed $BR(h\to \mu^+ \mu^-)$ is somewhat larger than $BR(h\to \tau^+ \tau^-)$ and both decays, purely arising  from new physics effects, can reach the level of ${\cal O}(10^{-3})$, with the SM predictions being at $2.18 \times 10^{-4}$ and $6.27 \times 10^{-2}$, respectively~\cite{LHCHiggsCrossSectionWorkingGroup:2016ypw}. When the SM contributions are included, the $\mu^+ \mu^-$ mode from the new effects is dominant while $\tau^+\tau^-$ can be maximally changed by $30\%$.  The reason that $h\to \tau^+\tau^-$ has a smaller BR is because the Yukawa couplings $y^{\ell}_{1k\tau}$ and $y^{\ell}_{2k\tau}$ obtain a stricter constraint from the $\tau \to \mu \gamma$ decay, whereas the Yukawa couplings $y^{\ell}_{2k\mu} y^\ell_{1k \mu}$ are only bounded by $\Delta a_\mu\in (1,5)\times 10^{-9}$.  The BRs for $h\to e \tau$ and $h\to \mu \tau$ as a correlation to  $BR(h\to \mu^+ \mu^-)$ are shown in Fig.~\ref{fig:Lepton_Higgs}(c) and (d), respectively. The resulting $BR(h\to \mu\tau)$ can reach ${\cal O}(10^{-4})$ while $BR(h\to e\tau)$ can be of ${\cal O}(10^{-5})$ in the model.  From the results in Fig.~\ref{fig:LFV_para}, it can be seen that  $|y^{\ell}_{1k\mu, 2k\mu}|$ have wide allowed-regions and a little bit larger values; therefore, it can be expected that the resulting $BR(h\to \mu\tau)$ can be larger than $BR(h\to e \tau)$.

\section{Summary} \label{sec:sum}

To explore possible mechanisms to enhance the top-FCNC and LFV processes, which are highly suppressed in the SM, we extend the SM by imposing a $Z_2$ symmetry and introducing an inert Higgs doublet, a singlet charged Higgs boson, and a singlet vector-like quark and neutral leptons.  It has been found that the enhancement of the top-FCNC decays has to rely on the Yukawa couplings of ${\cal O}(1)$.  Since the new Higgs trilinear couplings $h$-$H^\pm_I$-$\chi^\mp$ appear and their magnitudes can be much larger than the gauge couplings of electroweak interactions, the BR for the loop-induced $t\to c h$ can reach ${\cal O}(10^{-4})$ in the model and $t\to cZ$ is of ${\cal O}(10^{-5})$, though the BR for $t\to c\gamma$ stays below $10^{-6}$. 

To satisfy the strict limit from $\mu\to e\gamma$, we simply set $BR(\mu\to e \gamma)\approx 0$ to make the parameter scan more efficient. Based on the current upper limits of $\tau \to \ell \gamma$ and $h\to ee/e\mu$, it has been found that the BRs for $h\to \mu^+ \mu^-$ and $h\to \tau\tau$ can maximally reach ${\cal O}(10^{-3})$, with the former possibly having a somewhat larger BR.  In addition, the resulting $BR(h\to \mu \tau)$ of ${\cal O}(10^{-4})$ can be within the sensitivity of HL-LHC. 

The lepton $g-2$ generally can be either positive or negative in the model.  Using the constrained parameters, it is found that the predicted muon $g-2$ can match the experimental result within $2\sigma$ errors, where the hadronic vacuum polarization in the SM is obtained by the data-driven approach.  Moreover, because the involved Yukawa couplings are different from those in the muon $g-2$, the sign of electron $g-2$ in the model is not fixed.  Nevertheless, its magnitude can be consistent with the observed values in different atomic systems that are of opposite signs.

\section*{Acknowledgments}

This work was supported in part by the National Science and Technology Council, Taiwan under the Grant No.~MOST-110-2112-M-006-010-MY2 (C.~H.~Chen) and Grants No. MOST-108-2112-M-002-005-MY3 and No. 111-2112-M-002-018-MY3 (C.~W.~Chiang and C.~-W.~Su).

\end{document}